\documentclass[fleqn,usenatbib]{mnras}

\usepackage{newtxtext,newtxmath}
\usepackage[T1]{fontenc}
\usepackage{ae,aecompl}
\usepackage{graphicx}	
\usepackage{amsmath}	
\usepackage{amssymb}	
\usepackage{rotating}
\usepackage[usenames,dvipsnames]{color}
\usepackage{times}
\usepackage{epsfig}
\usepackage{natbib}
\usepackage{epstopdf} 
\usepackage{CJK}

\newcommand{\mbh}{$M_{\rm BH}$}

\newcommand{\s}{$\sigma_{\star}$}
\newcommand{\msun}{$M_{\odot}$}

\title[OIII width as surrogate for $\sigma_{\star}$?]{
 Studying the [OIII]$\lambda$5007\AA~emission-line width
  in a sample of $\sim$80 local active galaxies: A surrogate for $\sigma_{\star}$?} 

\author[V. N. Bennert et al.]{
  Vardha N. Bennert$^{1}$,
  Donald Loveland$^{1}$,
  Edward Donohue$^{1}$,
Maren Cosens$^{1,2}$
\newauthor
Sean Lewis$^{1,3}$,
S. Komossa$^{4}$,
Tommaso Treu$^{5}$,
Matthew A. Malkan$^{5}$,
\newauthor
Nathan Milgram$^{1}$,
Kelsi Flatland$^{1}$,
Matthew W. Auger$^{6}$,
Daeseong Park $^{7}$,
\newauthor
and Mariana S. Lazarova$^{8,9}$
\\
$^{1}$Physics Department, California Polytechnic State University, San Luis Obispo
CA 93407, USA; vbennert@calpoly.edu\\
$^{2}$Now at Center for Astrophysics and Space Sciences, University of
  California, San Diego, 9500 Gilman Drive, La Jolla, CA 92093, USA\\
$^{3}$Now at Department of Physics, Drexel University, Philadelphia, PA 19104, USA\\
  $^{4}$Max-Planck-Institut f{\"u}r Radioastronomie, Auf dem H{\"u}gel 69, 53121 Bonn, Germany\\
$^{5}$Department of Physics, University of California, Los Angeles, CA 90095, USA\\
$^{6}$Institute of Astronomy, Madingley Road, Cambridge CB3 0HA, UK\\
$^{7}$Korea Astronomy and Space Science Institute, Daejeon, 34055, Republic of Korea\\
$^{8}$Department of Physics and Astronomy, University of Nebraska Kearney, Kearney, NE 68849, USA\\
$^{9}$Now at Department of Physics and Astronomy, University of Northern Colorado, Greeley, CO 80639, USA
}

\date{Accepted 2018 August 10. Received 2018 July 30; in original form 2018 May 25.}

\pubyear{2018}

\begin{document}
\label{firstpage}
\pagerange{\pageref{firstpage}--\pageref{lastpage}}
\maketitle

\begin{abstract}
For a sample of $\sim$80 local ($0.02 \leq z \leq 0.1$) Seyfert-1 galaxies
with high-quality long-slit Keck spectra and spatially-resolved stellar-velocity
dispersion (\s) measurements, we study the profile of the 
[OIII]$\lambda$5007\AA~emission line to test the validity of using its width as a surrogate for \s.
Such an approach has often been used in the literature, since it is difficult to measure \s~for
type-1 active galactic nuclei (AGNs)
due to the AGN continuum outshining the stellar-absorption lines.
Fitting the [OIII] line with a single Gaussian or Gauss-Hermite polynomials
overestimates \s~by 50-100\%.
When line asymmetries from non-gravitational gas motion
are excluded in a double Gaussian fit,
the average ratio between the core [OIII] width ($\sigma_{\rm {[OIII],D}}$) and~\s~is $\sim$1,
but with individual data points off by up to a factor of two.
The resulting black-hole-mass-$\sigma_{\rm {[OIII],D}}$
relation scatters around that of quiescent galaxies and reverberation-mapped AGNs.
However, a direct comparison between \s~and
$\sigma_{\rm {[OIII],D}}$ shows no close correlation, only that
both quantities have the same range, average and standard deviation,
probably because they feel the same gravitational potential.
The large scatter is likely due to the fact that line profiles are a luminosity-weighted average,
dependent on the light distribution and underlying kinematic field.
Within the range probed by our sample (80-260\,km\,s$^{-1}$),
our results strongly caution against the use of [OIII] width as a surrogate for \s~on an individual basis.
Even though our sample consists of radio-quiet AGNs,
FIRST radio-detected objects have, on average, a $\sim$10\% larger [OIII] core width.
\end{abstract}


\section{Introduction}
\label{intro}
The relationship between the masses of supermassive black holes (BHs) and the properties of their host galaxies
has been amongst the most active research areas in contemporary astrophysics,
hinting at a co-evolution between BHs and galaxies
\citep[for a recent review see, e.g.,][]{kor13}.
Such a co-evolution can be explained either by
mutual growth via mergers
or by feedback from the active galactic nucleus (AGN)
in an evolutionary stage when the BH is growing through accretion.
AGNs are thus promising probes towards understanding the origin of these
BH mass (\mbh) scaling relations.
Unfortunately, the AGN emission (featureless non-stellar continuum
plus emission lines)
often outshines the host galaxy, making it
difficult to measure the host-galaxy  properties.
In particular, measuring
stellar-velocity dispersion ($\sigma_{\star}$),
which, of 
  all host galaxy properties, seems to scale the tightest with the BH mass
\citep{bei12, sha16}, is hampered by the contaminating AGN
continuum and emission lines.

To mitigate this problem,
several studies have suggested to
use the width of the [OIII]$\lambda$5007\AA~emission line
(hereafter [OIII]) originating in the narrow-line region (NLR)
as a surrogate for $\sigma_{\star}$,
 assuming that the NLR is gravitationally bound to the bulge
 and thus, that the gas kinematics follows the bulge potential
 \citep[e.g.,][]{ter90,whi92,nel96,nel00,shi03,bor03,gre05,net07,sal07,sal13}.
However, while the [OIII] emission line is a prominent line that can
 be easily measured in AGNs out to large distances,
 it is also known to often have asymmetric line profiles
 due to non-gravitational gas kinematics such as outflows, infalls, or interaction
 with radio jets. In particular, it is known to often display a blue wing
 \citep[e.g.,][]{hec81,der84,whi85,wil85,mul13,woo16},
 generally interpreted as a signature of outflows
 with dust preferentially hiding one cone behind the stellar disk.
 For that reason,
 some studies have excluded the [OIII] blue wing, as well as any  radio sources and
 galaxies undergoing tidal interactions.  The
 \mbh~was found to scale with the width of the [OIII] line ($\sigma_{\rm {[OIII]}}$),
 albeit with a large scatter \citep[e.g.,][]{nel96,gre05}.
 Other studies have suggested the use of  different 
 emission lines,
  such as [SII]$\lambda\lambda$6716, 6731 \citep[e.g.,][]{kom07,ho09} that have a lower ionization potential and do not suffer from substantial asymmetries,
  or mid-infrared lines \citep[e.g.,][]{das08,das11}, but the scatter
  is comparable to that of the core of the [OIII] line.
  While all studies confirm the original findings by \citet{nel96},
  i.e.~a moderately strong correlation between
  $\sigma_{\star}$ and $\sigma_{\rm {[OIII]}}$ but with real scatter,
  the origin of the scatter remains unclear. No dependencies
  have been found with AGN luminosity, host galaxy morphology,
  star formation rate, or local environment \citep{gre05,ric06}.

  However, unlike the original study by \citet{nel96},
very few previous studies have 
measured both properties,
$\sigma_{\star}$ and $\sigma_{\rm {[OIII]}}$, directly and simultaneously
for a given sample,
mainly due to the difficulties of measuring stellar-velocity dispersion
  in type-1 active galaxies. 
Often, 
conclusions are instead drawn by comparing
the \mbh-$\sigma_{\rm {[OIII]}}$ relation for type-1 galaxies
to the  \mbh-$\sigma_{\star}$ relation for quiescent galaxies
\citep{nel00,kom07},
or by comparing \mbh~derived from $\sigma_{\rm {[OIII]}}$ to
\mbh~derived from reverberation mapping \citep{nel00} or the virial method
using H$\beta$ \citep{bor03}.
\citet{bon05} predict $\sigma_{\star}$ indirectly from the Faber-Jackson relation
and conclude, from studying
the $M_{\rm host}$-$\sigma_{\rm {[OIII]}}$ relationship
for a sample of 21 radio-quiet quasars,
that $\sigma_{\rm {[OIII]}}$ is on average consistent with $\sigma_{\star}$.
Similarly, \citet{sal15} find agreement with the Faber-Jackson relation
when using the width of the [OIII] emission line as a proxy for stellar-velocity dispersion,
supporting the general utility of the [OIII] line width as a surrogate for $\sigma_{\star}$
in statistical studies.
\citet{gru04} and \citet{wan01} use  $\sigma_{\rm {[OIII]}}$ to
 investigate their \mbh~distributions of narrow-line Seyfert-1 galaxies (NLSy1).
\citet{gre05} compare $\sigma_{\rm {[OIII]}}$ to $\sigma_{\star}$ directly,
but for a sample of type-2 Seyfert galaxies.
Similarly, \citet{woo16} use a sample of 39,000 type-2 AGNs at $z<0.3$ from SDSS
and find a broad relation between [OIII] and $\sigma_{\star}$,
but with [OIII] being wider by 30-40\% since wings are not excluded
from the fit.
However, for a sub-sample of AGNs for which the [OIII] profile is well fitted  by
a single Gaussian model, \citet{woo16} find that the velocity dispersion
is comparable to the stellar-velocity dispersion.
\citet{ric06} use spatially-resolved HST/STIS spectra
for a sample of mostly type-2 Seyfert galaxies and find
that NLR line widths underestimate $\sigma_{\star}$.
Other studies have assumed that $\sigma_{\rm {[OIII]}}$
traces $\sigma_{\star}$ and used it to probe cosmic evolution
\citep{shi03, sal13}.
Also, most studies cited above use the width of the entire [OIII] emission
line, possibly including non-gravitational motion,
even though already \citet{nel96} showed that the [OIII] line profile base
and wings do not correlate as tightly with stellar-velocity dispersion
as the [OIII] core \citep[similar conclusions were also reached by][]{gre05}.

Thus, despite the widespread use of $\sigma_{\rm {[OIII]}}$ as a substitute for
$\sigma_{\star}$, caution is in order.

We have recently presented a baseline of the \mbh-$\sigma_{\star}$
relation for active galaxies
for a sample of 65 Seyfert-1 galaxies 
in the local Universe selected from the Sloan
Digital Sky Survey (SDSS) \citep[][]{ben15}.
SDSS images are used to determine host-galaxy morphology
and AGN luminosity free of host-galaxy contamination.
High signal-to-noise ratio Keck spectra yield H$\beta$ line width
to estimate \mbh~and spatially-resolved stellar-velocity dispersion \citep[][]{ben11a,har12}.
Thus, our sample is uniquely suited to study
the direct relationship between $\sigma_{\star}$ and $\sigma_{\rm {[OIII]}}$
for a homogeneous sample of local Seyfert-1 galaxies.
Moreover, we make use of the spatially-resolved Keck spectra to
isolate the nuclear line profile and to probe spatial dependencies.
We compare the resulting \mbh-$\sigma_{\rm {[OIII]}}$ relation
to the \mbh-$\sigma_{\star}$ relation \citep{ben15} and look
for trends with host galaxy and nuclear properties.

The paper is organized in the following manner.
Section~\ref{sample} summarizes the sample selection,
observations, and data reduction.  Section~\ref{analysis} describes the analysis
 of the data. Section~\ref{results} discusses
 the derived quantities and results.
 Section~\ref{summary} concludes with a summary.
 Note that the paper presents, first, a
 traditional approach focused on velocity dispersion ratios
 and their correlations to  \mbh,
 and then discusses the correlation between kinematic estimators directly
 and the shortcomings of conclusions based solely on ratios.
Throughout the paper, a Hubble constant of H$_{\rm{0}}$
 = 70 km s$^{-1}$, $\Omega_{\rm{\lambda}}$ = 0.7, and $\Omega_{\rm{M}}$ = 0.3 are assumed.

\section{Sample Selection, Observations, and Data Reduction}
\label{sample}
Sample selection, observations, and data reduction
are described in detail in previous papers, in which we are
focusing on the BH mass scaling relations for this sample
\citep{ben11a,har12,ben15}.
In brief, 102 type-1 Seyfert galaxies were
selected from the SDSS data release six (DR6) based on
redshift (0.02 $\leq z \leq$ 0.1) and \mbh~($>10^{7}M_{\odot}$).
They were observed with the Low Resolution Imaging Spectrometer
(LRIS) at the Keck 10-m telescope  between January 2009 and March 2010,
using a 1" wide, 175" long slit aligned with the major axis of the host galaxy
(as determined from SDSS images), with exposure times ranging from 600 to 1200 s. 
Here, we use only the blue spectra, covering a range of 
$\sim$3200-5350\AA \space and an instrumental resolution of 88 km\,s$^{-1}$ ($R \simeq 3000$).
The instrumental resolution of our aperture spectra was determined from
  the [OI] 5577\AA~atmospheric emission line as the square root of the second moment
  (which is approximately FWHM/2.355 for a Gaussian)
  and subtracted in quadrature from the width measurements.

Data were reduced following standard reduction steps
(bias subtraction, flat field correction, cosmic ray rejection, wavelength calibration,
and relative flux calibration).
Spatially-resolved spectra were extracted at the center of each galaxy
and offset in either direction along the major axis \citep[see][for more details]{har12}.
We make use of these spatially-resolved spectra to compare $\sigma_{\star}$
\citep{har12} to the [OIII] line-width ($\sigma_{\rm {[OIII]}}$) at different distances
from the nucleus. For each galaxy, we extracted the central spectrum
plus up to five spectra on either side of the center (out to 5"), giving a total of 11 spectra.
However, not all spectra were used for all galaxies, depending on the S/N,
available $\sigma_{\star}$ measurement in \citet{har12} and presence of the [OIII] emission line.
Additionally, we also use aperture spectra within the bulge effective
radius, as determined in \citet{ben15}, resulting in one additional spectrum per galaxy.
We also included the [OII]$\lambda$3727\AA~line (hereafter [OII]) in this comparison.
However, given that the [OII] line is much weaker than the [OIII] line in these AGN-powered spectra,
we can only fit the [OII] line for the central row, as well as within the effective radius.
Our final sample consists of 81 galaxies for which we have at least one
$\sigma_{\rm {[OIII]}}$ measurement.

\section{Analysis}
\label{analysis}

\subsection{Fits to [OIII]}
To fit the emission lines around [OIII], a multi-component spectral decomposition code
is used (described in detail in \citet{par15}).
The continuum is modeled by a combination of
AGN featureless non-stellar continuum, AGN Fe II emission template \citep{bor92}, and
host galaxy starlight templates
from the Indo-US spectral library \citep{val04}.
The broad H$\beta$ emission line is fitted by 
Gauss-Hermite polynomials (order 3-6)
\citep[][]{mar93, woo06, mcg08}.
The [OIII]$\lambda\lambda$4959,5007\AA~emission lines
are fitted keeping their flux ratio fixed at 1:3.
The [OIII]$\lambda$5007\AA~fit is used as a template for the narrow H$\beta$,
with the flux ratio as a free parameter.
For examples of fits to the central spectra for the entire region around H$\beta$, see \citet{ben15}.

Three different approaches are used to fit [OIII]$\lambda$5007\AA:
(1) a single Gaussian is fitted, with the
resulting width being referred to in the following as $\sigma_{\rm [OIII],S}$;
(2) a double Gaussian is
fitted,
with the resulting width of the central component only being referred to in the following as
$\sigma_{\rm [OIII],D}$;
(3) a Gauss-Hermite polynomial series (order 7-12) is fitted,
with the second moment of the full distribution (i.e., the line dispersion)
referred to as $\sigma_{\rm [OIII],GH}$.
The reasoning for the choice of these three fits is as follows.

If the cause for the line broadening is Doppler motion of the line
emitting gas, a Gaussian profile is expected.
While a single Gaussian can yield a reasonable fit in cases without line
asymmetries, asymmetries are known to occur especially for the [OIII] emission line
\citep[e.g.,][]{hec81,der84,whi85,wil85}.
Gauss-Hermite polynomials
can give the best fit to the overall line profile.
However, in case of asymmetries, we expect both the single Gaussian as well
as Gauss-Hermite polynomials to overestimate the width of the central [OIII] component.
This core component is the one we are interested in
since it is the one emitted from gas most likely to follow the
gravitational potential of the bulge.
To isolate this component
from gas motion, such as outflows and infalls reflected in blue or red wings,
we use a double Gaussian fit.
In some objects, the second Gaussian is used to fit an underlying broader central component,
indicating turbulent motion  \citep{kol13}.
When comparing the derived width to the
stellar-velocity dispersion, we only consider the Gaussian fitting of the central core
component, i.e.~the Gaussian with the higher peak and smaller width.

A single Gaussian is fitted for a total of 346 spectral rows,
a Gauss-Hermite polynomial for 336 spectral rows
and a double Gaussian for 326 spectral rows.
Note that, in cases of low S/N, fitting the line with a double Gaussian
can result in the wing component fitting noise. We thus carefully inspected
all fits by eye and excluded those cases. It is generally recommended
to only fit with a double Gaussian in cases of clear evidence of a broader wing component
and/or to enforce a peak-to-noise level of the second component of at least 3 \citep[see also,][]{woo16}.
In addition to S/N, spectral resolution is also important when fitting a double Gaussian.
A resolution much smaller than $R \simeq 3000$, as  is used here, would make this approach challenging.

\begin{figure*}
\includegraphics[width=\linewidth]{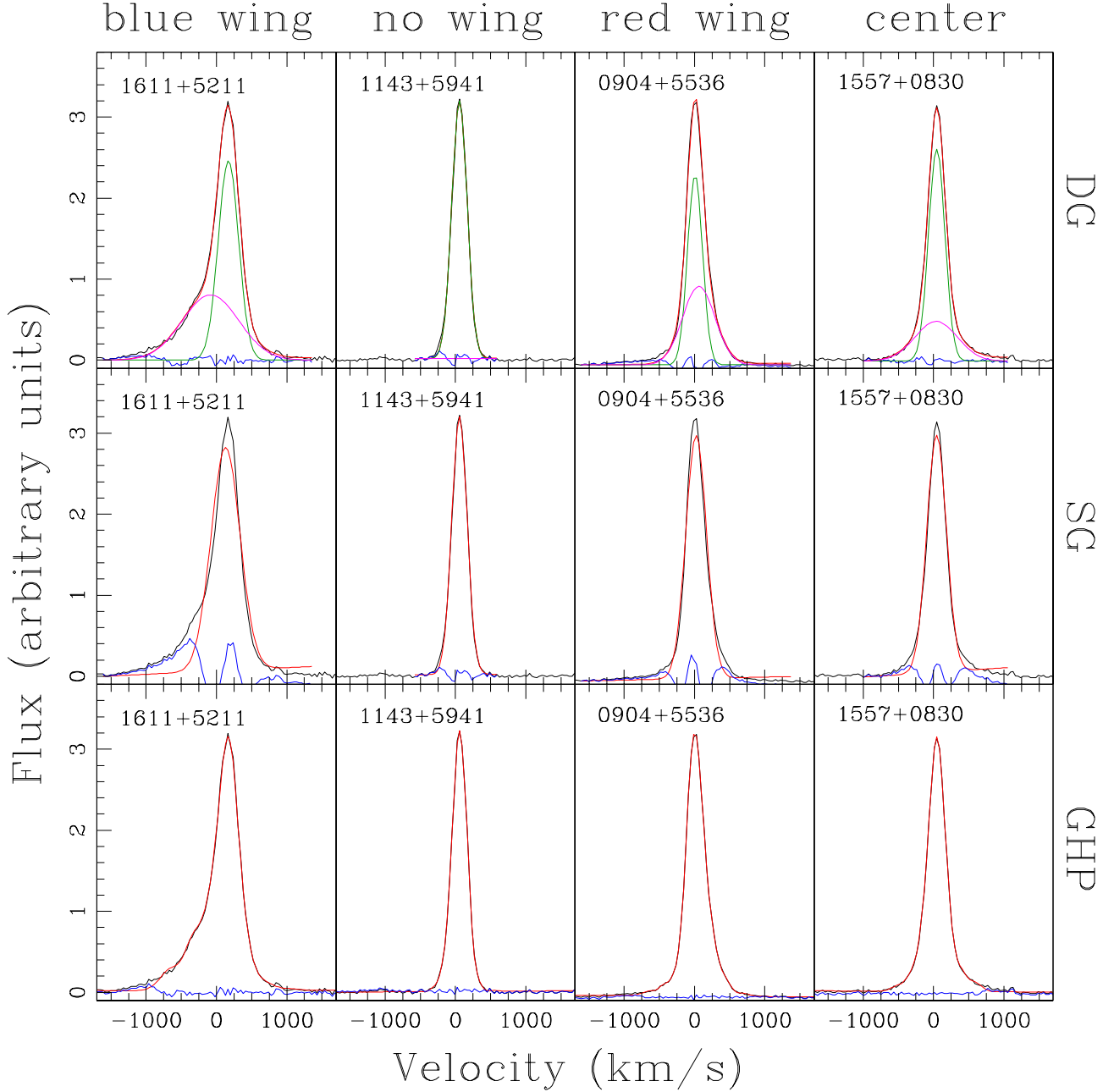}
\caption{Illustration of [OIII] fitting methods.
  Three objects are shown: in the first column, the [OIII] emission line
  of object 1611+5211 has a blue wing;
  in the second column, the [OIII] emission line of object 1143+5941 has no distinct wing;
  in the third column, the [OIII] emission line of object 0904+5536 has a red wing;
  in the fourth column, the [OIII] emission line of object 1557+0830 has a broader central component.
  The different [OIII] emission lines are fitted by a double Gaussian (DG; upper panels),
  a single Gaussian (SG; middle panels) and a Gauss-Hermite polynomial function (GHP; lower panels).
  The observed spectrum is shown in black, the total fit is in red. For the double Gaussian fit,
  the green line shows one Gaussian fitted to the central core [OIII] emission line,
  the magenta line
  shows one Gaussian fitted to the blue/red wing or broader central component.
  While a single Gaussian is only a good fit for lines without asymmetries
  (such as 1143+5941),
  the Gauss-Hermite polynomials give the best overall fit to the line.
  However, both the single Gaussian and the Gauss-Hermite polynomial
  fits overestimate the line width of
  the central core [OIII] emission line.}
\label{figure:oiiifits}
\end{figure*}

Figure~\ref{figure:oiiifits} illustrates our approach.
Tables~\ref{table:sample1}-\ref{table:spat} list the results.

\begin{table*}
\caption{Sample and quantities within effective bulge radius. Col. (1): target ID used throughout the text (based on R.A. and declination). 		      	   	  
Col. (2): Right ascension. 				  
Col. (3): Declination. 
Col. (4): Redshift from SDSS-DR7.
Col. (5): Logarithm of BH mass (solar units) (uncertainty of 0.4 dex).
Col. (6): Spheroid effective radius in kpc.
Col. (7): Stellar-velocity dispersion within
spheroid effective radius determined from CaH\&K (uncertainty of 0.04 dex).
Col. (8): [OIII] width within
spheroid effective radius determined from double Gaussian fit (central line only; uncertainty of 0.04 dex).
Col. (9): [OIII] width within
spheroid effective radius determined from single Gaussian fit (uncertainty of 0.04 dex).
Col. (10): [OIII] width within
spheroid effective radius determined from Gauss-Hermite polynomial fit (uncertainty of 0.04 dex).
Col. (11): [OII] width within spheroid effective radius (uncertainty of 0.04 dex).
Note that objects with ``no data'' in some of the columns are not included in \citet{ben15} since one of the quantities for the BH mass - $\sigma_{\star}$ relationship
could not be determined, but at least some spatially-resolved $\sigma_{\star}$ measurements exist in \citet{har12} and these objects are included
in the spatially-resolved [OIII] measurements in Table~\ref{table:spat}.
Col. (12): FIRST integrated radio flux.
ND = not detected.
NC = not covered, that is outside of the survey area.
}
\label{table:sample1}
\begin{tabular}{lccccccccccc}
  \hline
Object & 
R.A. &
Decl. &
$z$ &
log \mbh/\msun &
$r_{\rm eff, sph}$ &
$\sigma_{\rm \star}$ &
$\sigma_{\rm [OIII], D}$ &
$\sigma_{\rm [OIII], S}$ &
$\sigma_{\rm [OIII], GH}$ &
$\sigma_{\rm [OII]}$ &
FIRST\\
& (J2000) & (J2000) & & & (kpc)
& (km\,s$^{-1}$)
& (km\,s$^{-1}$)
& (km\,s$^{-1}$) 
& (km\,s$^{-1}$) 
& (km\,s$^{-1}$)
& (mJy)\\
(1) & (2)  & (3) & (4) &
(5) & (6) & (7) & (8) & (9) & (10) & (11) & (12)\\
\hline
0013$-$0951 & 00 13 35.38 & $-$09 51 20.9 & 0.0615 & 7.85 &  4.8 &   96 &  123 &  212 &  261 &  151 & ND\\ 
0026+0009 & 00 26 21.29 & +00 09 14.9 & 0.0600 & 7.05 &  1.8 &  172 &  190 &  190 &  193 &  171 & ND\\ 
0038+0034 & 00 38 47.96 & +00 34 57.5 & 0.0805 & 8.23 &  1.9 &  127 &  174 &  212 &  249 &  160 & 1.67\\ 
0109+0059 & 01 09 39.01 & +00 59 50.4 & 0.0928 & 7.52 &  0.3 &  183 &  144 &  263 &  310 &  175 & 1.09\\ 
0121$-$0102 & 01 21 59.81 & $-$01 02 24.4 & 0.0540 & 7.75 &  1.8 &   90 &  152 &  247 &  290 &  205 & 4.00\\ 
0150+0057 & 01 50 16.43 & +00 57 01.9 & 0.0847 & 7.25 &  4.5 &  176 &  131 &  174 &  245 &  149 & ND\\ 
0206$-$0017 & 02 06 15.98 & $-$00 17 29.1 & 0.0430 & 8.00 &  6.2 &  225 &  183 &  229 &  307 &  165 & ND\\ 
0212+1406 & 02 12 57.59 & +14 06 10.0 & 0.0618 & 7.32 &  1.0 &  171 &  152 &  181 &  223 &  158 & NC\\ 
0301+0110 & 03 01 24.26 & +01 10 22.8 & 0.0715 & ... & ... & ... & ... & ... & ... & ... & ND\\ 
0301+0115 & 03 01 44.19 & +01 15 30.8 & 0.0747 & 7.55 &  2.7 &   99 &  144 &  312 &  375 &  114 & ND\\ 
0336$-$0706 & 03 36 02.09 & $-$07 06 17.1 & 0.0970 & 7.53 & 12.9 &  236 &  138 &  188 &  230 &  238 & ND\\ 
0353$-$0623 & 03 53 01.02 & $-$06 23 26.3 & 0.0760 & 7.50 &  1.6 &  175 &  113 &  155 &  177 &  131 & ND\\ 
0735+3752 & 07 35 21.19 & +37 52 01.9 & 0.0962 & ... & ... & ... & ... & ... & ... & ... & ND\\ 
0737+4244 & 07 37 03.28 & +42 44 14.6 & 0.0882 & 7.55 &  4.2 & ... & ... & ... & ... & ... & 1.01\\ 
0802+3104 & 08 02 43.40 & +31 04 03.3 & 0.0409 & 7.43 &  2.8 &  116 & ... & ... & ... & ... & ND\\ 
0811+1739 & 08 11 10.28 & +17 39 43.9 & 0.0649 & 7.17 &  2.5 &  142 &  103 &  124 &  138 &  111 & ND\\ 
0813+4608 & 08 13 19.34 & +46 08 49.5 & 0.0540 & 7.14 &  1.0 &  122 &  100 &  116 &  145 &  109 & ND\\ 
0831+0521 & 08 31 07.62 & +05 21 05.9 & 0.0635 & ... & ... & ... & ... & ... & ... & ... & 1.73\\ 
0845+3409 & 08 45 56.67 & +34 09 36.3 & 0.0655 & 7.37 &  1.4 &  123 &   89 &  121 &  179 &  103 & ND\\ 
0857+0528 & 08 57 37.77 & +05 28 21.3 & 0.0586 & 7.42 &  2.5 &  126 &  124 &  156 &  194 &  124 & ND\\ 
0904+5536 & 09 04 36.95 & +55 36 02.5 & 0.0371 & 7.77 &  4.0 &  194 &  144 &  173 &  216 &  155 & 1.35\\ 
0909+1330 & 09 09 02.35 & +13 30 19.4 & 0.0506 & ... & ... & ... & ... & ... & ... & ... & ND\\ 
0921+1017 & 09 21 15.55 & +10 17 40.9 & 0.0392 & 7.45 &  2.6 & ... &  109 &  161 &  211 &  109 & ND\\ 
0923+2254 & 09 23 43.00 & +22 54 32.7 & 0.0332 & 7.69 &  0.9 &  149 &  158 &  275 &  316 &  285 & 9.29\\ 
0923+2946 & 09 23 19.73 & +29 46 09.1 & 0.0625 & 7.56 &  4.2 &  142 &  102 &  117 &  151 &  119 & ND\\ 
0927+2301 & 09 27 18.51 & +23 01 12.3 & 0.0262 & 6.94 &  7.1 &  196 &  172 &  198 &  241 &  185 & 2.79\\ 
0932+0233 & 09 32 40.55 & +02 33 32.6 & 0.0567 & 7.44 &  0.7 &  126 &  121 &  152 &  169 &  131 & ND\\ 
0932+0405 & 09 32 59.60 & +04 05 06.0 & 0.0590 & ... & ... & ... & ... & ... & ... & ... & ND\\ 
0938+0743 & 09 38 12.27 & +07 43 40.0 & 0.0218 & ... & ... & ... & ... & ... & ... & ... & ND\\ 
0948+4030 & 09 48 38.43 & +40 30 43.5 & 0.0469 & ... & ... & ... & ... & ... & ... & ... & ND\\ 
1002+2648 & 10 02 18.79 & +26 48 05.7 & 0.0517 & ... & ... & ... & ... & ... & ... & ... & ND\\ 
1029+1408 & 10 29 25.73 & +14 08 23.2 & 0.0608 & 7.86 &  3.0 &  185 &  163 &  182 &  224 &  179 & 1.33\\ 
1029+2728 & 10 29 01.63 & +27 28 51.2 & 0.0377 & 6.92 &  2.6 &  112 &  133 &  169 &  213 &  142 & ND\\ 
1029+4019 & 10 29 46.80 & +40 19 13.8 & 0.0672 & 7.68 &  2.0 &  166 &  170 &  210 &  266 &  168 & ND\\ 
1042+0414 & 10 42 52.94 & +04 14 41.1 & 0.0524 & 7.14 &  3.2 & ... &  133 &  157 &  207 &  135 & ND\\ 
1049+2451 & 10 49 25.39 & +24 51 23.7 & 0.0550 & 8.03 &  1.3 &  162 &  141 &  161 &  209 &  160 & ND\\ 
1058+5259 & 10 58 28.76 & +52 59 29.0 & 0.0676 & 7.50 &  1.3 &  122 &  116 &  152 &  187 &  145 & ND\\ 
1101+1102 & 11 01 01.78 & +11 02 48.8 & 0.0355 & 8.11 &  5.8 &  197 &  161 &  224 &  253 &  232 & 2.86\\ 
1104+4334 & 11 04 56.03 & +43 34 09.1 & 0.0493 & 7.04 &  1.1 & ... &  108 &  155 &  203 &  127 & ND\\ 
1116+4123 & 11 16 07.65 & +41 23 53.2 & 0.0210 & 7.23 &  1.6 &  108 &  149 &  174 &  252 &  162 & 2.27\\ 
1118+2827 & 11 18 53.02 & +28 27 57.6 & 0.0599 & ... & ... & ... & ... & ... & ... & ... & ND\\ 
1137+4826 & 11 37 04.17 & +48 26 59.2 & 0.0541 & 6.74 &  1.1 &  155 &  152 &  241 &  257 &  175 & 2.71\\ 
1140+2307 & 11 40 54.09 & +23 07 44.4 & 0.0348 & ... & ... & ... & ... & ... & ... & ... & ND\\ 
1143+5941 & 11 43 44.30 & +59 41 12.4 & 0.0629 & 7.51 &  3.8 &  122 &  111 &  119 &  150 &  124 & ND\\ 
1144+3653 & 11 44 29.88 & +36 53 08.5 & 0.0380 & 7.73 &  1.0 &  168 &  120 &  190 &  229 &  151 & ND\\ 
1145+5547 & 11 45 45.18 & +55 47 59.6 & 0.0534 & 7.22 &  1.4 &  118 &  136 &  201 &  241 &  156 & ND\\ 
1147+0902 & 11 47 55.08 & +09 02 28.8 & 0.0688 & 8.39 &  3.4 &  147 &  151 &  175 &  204 &  178 & 1.15\\ 
1205+4959 & 12 05 56.01 & +49 59 56.4 & 0.0630 & 8.00 &  2.4 &  152 &  175 &  217 &  244 &  202 & 1.79\\ 
1206+4244 & 12 06 26.29 & +42 44 26.1 & 0.0520 & ... & ... & ... & ... & ... & ... & ... & ND\\ 
1210+3820 & 12 10 44.27 & +38 20 10.3 & 0.0229 & 7.80 &  0.6 &  141 &  133 &  179 &  200 &  150 & 5.88\\ 
1223+0240 & 12 23 24.14 & +02 40 44.4 & 0.0235 & 7.10 &  3.4 &  124 &  120 &  170 &  198 &  181 & ND\\ 
1228+0951 & 12 28 11.41 & +09 51 26.7 & 0.0640 & ... & ... & ... & ... & ... & ... & ... & ND\\ 
1231+4504 & 12 31 52.04 & +45 04 42.9 & 0.0621 & 7.32 &  1.5 &  169 &  205 &  306 &  417 &  229 & 5.56\\ 
1241+3722 & 12 41 29.42 & +37 22 01.9 & 0.0633 & 7.38 &  1.7 &  144 &  132 &  174 &  212 &  184 & ND\\ 
\hline
\end{tabular}
\end{table*}

\begin{table*}
\caption{Table~\ref{table:sample1} continued.
}
\label{table:sample2}
\begin{tabular}{lccccccccccc}
  \hline
Object & 
R.A. &
Decl. &
$z$ &
log \mbh/\msun &
$r_{\rm eff, sph}$ &
$\sigma_{\rm \star}$ &
$\sigma_{\rm [OIII], D}$ &
$\sigma_{\rm [OIII], S}$ &
$\sigma_{\rm [OIII], GH}$ &
$\sigma_{\rm [OII]}$ &
FIRST\\
& (J2000) & (J2000) & & & (kpc)
& (km\,s$^{-1}$)
& (km\,s$^{-1}$)
& (km\,s$^{-1}$) 
& (km\,s$^{-1}$) 
& (km\,s$^{-1}$)
& (mJy)\\
(1) & (2)  & (3) & (4) &
(5) & (6) & (7) & (8) & (9) & (10) & (11) & (12)\\
\hline
1246+5134 & 12 46 38.74 & +51 34 55.9 & 0.0668 & 6.93 &  3.9 &  119 &  116 &  132 &  162 &  148 & ND\\ 
1250$-$0249 & 12 50 42.44 & $-$02 49 31.5 & 0.0470 & ... & ... & ... & ... & ... & ... & ... & ND\\ 
1306+4552 & 13 06 19.83 & +45 52 24.2 & 0.0507 & 7.16 &  2.3 &  114 &  122 &  161 &  212 &  117 & ND\\ 
1312+2628 & 13 12 59.59 & +26 28 24.0 & 0.0604 & 7.51 &  1.7 &  109 &  103 &  126 &  217 &  113 & ND\\ 
1313+3653 & 13 13 48.96 & +36 53 57.9 & 0.0667 & ... & ... & ... & ... & ... & ... & ... & ND\\ 
1323+2701 & 13 23 10.39 & +27 01 40.4 & 0.0559 & 7.45 &  0.9 &  124 &  158 &  219 &  276 &  184 & ND\\ 
1353+3951 & 13 53 45.93 & +39 51 01.6 & 0.0626 & ... & ... & ... & ... & ... & ... & ... & ND\\ 
1405$-$0259 & 14 05 14.86 & $-$02 59 01.2 & 0.0541 & 7.04 &  0.6 &  125 &  132 &  189 &  235 &  106 & ND\\ 
1416+0137 & 14 16 30.82 & +01 37 07.9 & 0.0538 & 7.26 &  3.6 &  173 &  182 &  291 &  342 &  211 & 1.70\\ 
1419+0754 & 14 19 08.30 & +07 54 49.6 & 0.0558 & 8.00 &  5.4 &  215 &  211 &  285 &  354 &  187 & 4.49\\ 
1423+2720 & 14 23 38.43 & +27 20 09.7 & 0.0639 & ... & ... & ... & ... & ... & ... & ... & ND\\ 
1434+4839 & 14 34 52.45 & +48 39 42.8 & 0.0365 & 7.66 &  0.9 &  109 &  132 &  178 &  207 &  150 & ND\\ 
1535+5754 & 15 35 52.40 & +57 54 09.3 & 0.0304 & 8.04 &  2.8 &  110 &  175 &  208 &  244 &  147 & 5.32\\ 
1543+3631 & 15 43 51.49 & +36 31 36.7 & 0.0672 & 7.73 &  3.8 &  146 &  140 &  221 &  257 &  218 & ND\\ 
1545+1709 & 15 45 07.53 & +17 09 51.1 & 0.0481 & 8.03 &  1.1 &  163 &  143 &  172 &  225 &  153 & ND\\ 
1554+3238 & 15 54 17.42 & +32 38 37.6 & 0.0483 & 7.87 &  1.7 &  158 &  200 &  235 &  287 &  217 & 2.52\\ 
1605+3305 & 16 05 02.46 & +33 05 44.8 & 0.0532 & 7.82 &  1.6 &  187 &  122 &  128 &  137 &  120 & ND\\ 
1606+3324 & 16 06 55.94 & +33 24 00.3 & 0.0585 & 7.54 &  1.7 &  157 &  192 &  226 &  263 &  155 & ND\\ 
1611+5211 & 16 11 56.30 & +52 11 16.8 & 0.0409 & 7.67 &  1.3 &  116 &  152 &  259 &  349 &  182 & 3.67\\ 
1636+4202 & 16 36 31.28 & +42 02 42.5 & 0.0610 & 7.86 &  9.7 &  205 &  197 &  227 &  314 &  130 & 1.18\\ 
1655+2014 & 16 55 14.21 & +20 14 42.0 & 0.0841 & ... & ... & ... & ... & ... & ... & ... & ND\\ 
1708+2153 & 17 08 59.15 & +21 53 08.1 & 0.0722 & 8.20 &  8.1 &  231 &  182 &  238 &  306 &  405 & ND\\ 
2221$-$0906 & 22 21 10.83 & $-$09 06 22.0 & 0.0912 & 7.77 &  6.1 &  142 &  126 &  154 &  198 &  134 & ND\\ 
2222$-$0819 & 22 22 46.61 & $-$08 19 43.9 & 0.0821 & 7.66 &  1.7 &  122 &  208 &  464 &  500 &  209 & 4.22\\ 
2233+1312 & 22 33 38.42 & +13 12 43.5 & 0.0934 & 8.11 &  2.1 &  193 &  160 &  257 &  303 &  196 & NC\\ 
2327+1524 & 23 27 21.97 & +15 24 37.4 & 0.0458 & 7.52 &  6.6 &  225 &  133 &  271 &  335 &  173 & NC\\ 
2351+1552 & 23 51 28.75 & +15 52 59.1 & 0.0963 & 8.08 &  2.5 &  186 &  101 &  232 &  245 &  254 & NC\\ 
\hline
\end{tabular}
\end{table*}

\begin{table*}
  \caption{Spatially-resolved quantities. Col. (1): Target ID used throughout the text (based on R.A. and declination).
    Col. (2): Offset of spatially resolved spectrum from center.
Col. (3): Spatially resolved stellar-velocity dispersion determined from CaH\&K (uncertainty of 0.04 dex) taken from \citet{har12}.
Col. (4): Spatially resolved [OIII] width determined from double Gaussian fit (central line only; uncertainty of 0.04 dex).
Col. (5): Spatially resolved [OIII] width determined from single Gaussian fit (uncertainty of 0.04 dex).
Col. (6): Spatially resolved [OIII] width determined from Gauss-Hermite polynomial fit (uncertainty of 0.04 dex).
This table is available in its entirety in machine-readable form in the online journal. A portion is shown here for guidance
regarding its form and content.
}
\label{table:spat}
\begin{tabular}{lccccc}
  \hline
Object & 
Offset &
$\sigma_{\rm \star}$ &
$\sigma_{\rm [OIII], D}$ &
$\sigma_{\rm [OIII], S}$ &
$\sigma_{\rm [OIII], GH}$\\
& arcsec
& (km\,s$^{-1}$) 
& (km\,s$^{-1}$)
& (km\,s$^{-1}$) 
& (km\,s$^{-1}$) \\
(1) & (2)  & (3) & (4) &
(5) & (6)\\
\hline
0013$$$-$$$0951 & +0.00       & 113 & 131 & 222 & 299\\ 
0013$$$-$$$0951 & +0.68       & 135 & 131 & 215 & 281\\ 
0013$$$-$$$0951 & +1.62       & ... & 313 & 378 & 520\\
0013$$$-$$$0951 & $$$-$$$0.68 & 119 & 169 & 236 & 340\\
0013$$$-$$$0951 & $$$-$$$1.62 & 162 & 386 & 392 & 612\\
\hline
\end{tabular}
\end{table*}

\subsection{Fits to [OII]}
The [OII]$\lambda$3727\AA~emission line is really a blended doublet line of
[OII]$\lambda$3726,3729\AA\AA.
It is a line with a lower ionization potential
(13.6 eV compared to 35 eV for [OIII]), emitted at larger distances from the
nucleus and, as such, spectra are expected to be less complex and dominated
by rotation. ([OIII] emitted from closer in can be more affected
by outflows and winds from the accretion disk, e.g.)
We thus fitted the line with a double Gaussian centered on the doublet, forcing
both lines to have the same width, but leaving the ratio as a free parameter
since it depends on electron density.
We used the resulting width (of a single Gaussian) as $\sigma_{\rm [OII]}$.
However, the [OII] line is weaker than the [OIII] line
and can only be fitted for the central row as well as within the effective radius.

  \subsection{Stellar-velocity dispersion}
  Spatially-resolved stellar-velocity dispersion measurements were taken from
  \citet{ben11a} and \citet{har12}, stellar-velocity dispersion measurements within the
  bulge effective radius (determined from surface photometry fitting of SDSS
  images) from \citet[][their equation (1)]{ben15}. For details, including examples of the fits,
  we refer the reader to those papers.
  In short, $\sigma_{\star}$ was measured from three different
  spectral regions, around CaH\&K$\lambda$$\lambda$3969, 3934\AA~(hereafter
CaH\&K), around the Mg Ib$\lambda$$\lambda$$\lambda$5167, 5173, 5184\AA~(hereafter
MgIb) lines and around Ca II$\lambda$$\lambda$$\lambda$8498, 8542, 8662\AA~(hereafter
CaT), fitting a linear combination of Gaussian-broadened
template spectra (G and K giants of various temperatures as well as spectra of A0 and F2 giants from the
Indo-US survey) and a polynomial continuum using a Markov chain Monte Carlo (MCMC) routine, following \citet{vdm94}. We used the resulting $\sigma_{\star}$ from the CaT region,
if available, else from CaH\&K and finally from MgIb,
if the two former were not available.

\section{Results and Discussion}
\label{results}
We here compare the resulting widths for [OIII] and [OII] with
$\sigma_{\star}$.
All 81 objects have at least one $\sigma_{\rm {[OIII]}}$ measurement.
Quantities necessary for comparison of
$\sigma_{\rm {[OIII]}}$ and $\sigma_{\star}$ for aperture spectra
within the effective bulge radius
are available for 62 of the 81 objects and, thus, the
\mbh-$\sigma_{\rm {[OIII]}}$ relation
is compared directly to the
M$_{\rm{BH}}$-$\sigma_{\star}$  relation for that sub-sample of 62 objects \citep{ben15}.
Likewise, when including $\sigma_{\rm [OII]}$ within the effective radius in the comparison,
a total of 62 objects are compared.

\subsection{[OIII] profile}
The double Gaussian fit reveals information on the general [OIII] line profile.
For 66\% of objects/spectral rows, the double Gaussian fitting resulted
in the fitting of a blue wing (-500\,km\,s$^{-1}$ $\le$ $v$ $\le$ -25\,km\,s$^{-1}$).
For 22\% of objects/spectral rows, a
Gaussian redshifted compared to the central core was fitted, implying a red wing
(25\,km\,s$^{-1}$ $\le$ $v$ $\le$ 500\,km\,s$^{-1}$).
For 12\% of objects/rows, the second Gaussian fitted a broader central component
(-25\,km\,s$^{-1}$ $\le$ $v$ $\le$ 25\,km\,s$^{-1}$).

\begin{figure}
  \includegraphics[width=\linewidth]{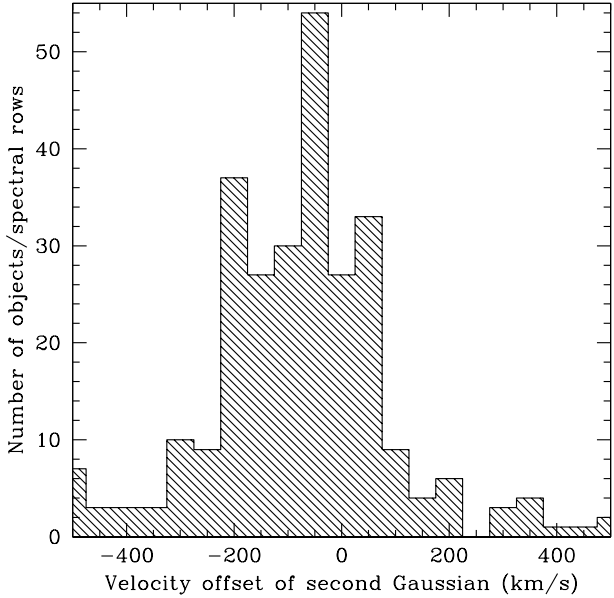}
  \caption{
Histogram of the velocity offset of the second Gaussian (the wing component) compared
to the central core Gaussian for all objects and spectral rows.}
\label{figure:oiiihisto}
\end{figure}

\begin{figure}
\includegraphics[width=\columnwidth]{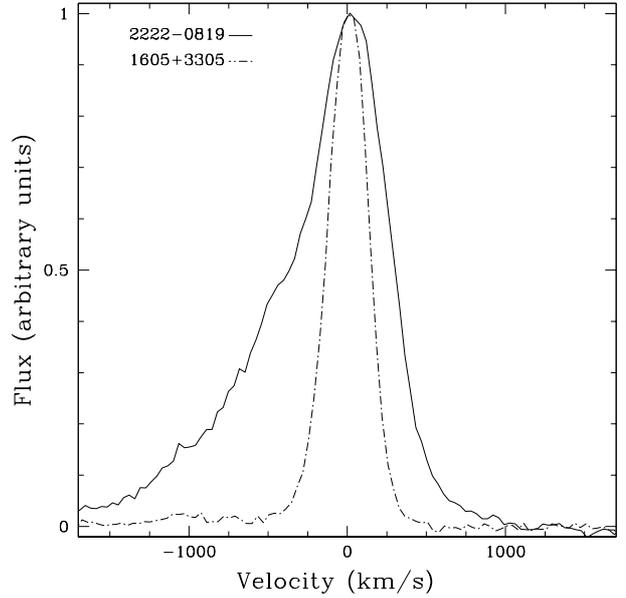}
  \caption{
      Examples of central [OIII] profiles for
      broadest [OIII] line (2222-0819,
      $\sigma_{\rm [OIII], GH}$ = 514 km\,s$^{-1}$) 
    and narrowest [OIII] line (1605+3305, $\sigma_{\rm [OIII], GH}$ = 127 km\,s$^{-1}$).
    For comparison, the local continuum was
    subtracted and the peak flux scaled to 1.
 \label{figure:oiiibroadnarrow}}
\end{figure}

The histogram of the velocity offset of the second Gaussian (the wing component) compared
to the central core Gaussian is shown in Figure~\ref{figure:oiiihisto}, including
all objects and spectral rows.
The average velocity offset for the blue wing is -155$\pm$7\,km\,s$^{-1}$,
and for the red wing 124$\pm$13\,s$^{-1}$, respectively.
While these results are overall comparable with those of \citet{woo16} for a sample
of $\sim$39,000 type-2 AGNs in SDSS, we find an even higher fraction of kinematic
signatures for outflows, likely because of the type-1 nature of our objects
for which the viewing angle is favorable to see outflows.
Indeed, the average [OIII] profile for type-1 AGNs, as determined from a sample of $\sim$10,000 AGNs from SDSS,
shows a strong blue wing that can be well fitted by a
broad second Gaussian component (average velocity offset of -148\,km\,s$^{-1}$)
\citep{mul13}.
Figure~\ref{figure:oiiibroadnarrow} shows examples of
the broadest and the narrowest [OIII] emission line profile.

\begin{figure}
\includegraphics[width=\linewidth]{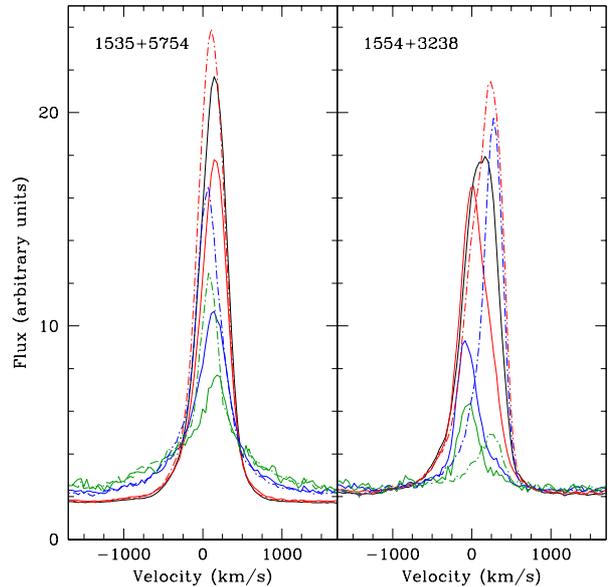}
  \caption{
 Examples of two objects that show a spatially changing [OIII] emission line profile
  (center = black; ``negative'' offset from center = solid lines; ``positive'' offset from center = dash-dotted lines;
red = 0.68'' distance; blue = 1.62'' distance; green = 2.84'' distance).  
Left: 1535+5754, observed at a position angle (p.a.) of 100deg,
with broader lines further out from the center (``negative''=south-east).
This galaxy does not show a strong rotation curve
\citep{ben11a}.
      Right: 1554+3238, observed at a p.a. of 80deg. In addition to the rotation curve
      ($\pm$200km/s) also visible from the stellar-absorption lines \citep{ben11a},
      the object shows a blue wing on the ``positive'' side of the center (south-east)
      and a red wing on the ``negative''
      side (north-west).}
  \label{figure:oiiichange}
\end{figure}

The [OIII] line shows rotation in at least 17\% of objects
with rotational velocities up to $\sim$$\pm$250\,km\,s$^{-1}$,
matching those of the stellar rotation curve \citep{har12}.
In 15\% of objects do we see evidence for HII regions in the outer spectra,
as traced by a sudden peak in [OIII] along with an increase in the H$\beta$/[OIII] ratio.
However, since H$\alpha$ is not covered by our spectra, we cannot
verify the origin of the ionization of these regions and thus do not further discuss them here.
There is a small fraction of objects ($\sim$7\%) that shows evidence
for a change in the [OIII] profile as a function of distance from the center,
with the majority showing a red wing on one side of the galaxy center
and a blue wing on the other, and some galaxies with the blue wing only present on one
side of the galaxy center (Figure~\ref{figure:oiiichange}).

Other than that, we do not find any trends with distance
from the center. For example, the ratio of broad (wing) [OIII] to narrow (core) [OIII]
does not change significantly as a function of radius (when fitted by a double Gaussian);
nor does the width of the broad [OIII] component change with radius.
Part of this is likely due to the fact that (i) the spectra are restricted
to the central few kpc, given the S/N ratio, and (ii) that the central
1-2 kpc are unresolved due to the ground-based seeing.
(The 1" width of the long-slit was chosen to match the seeing.
  1" corresponds to 0.43kpc for the smallest redshift of z=0.021 of our sample,
  to 1.8kpc for the largest redshift of z=0.097, and to 1.1kpc for the average redshift of z=0.058.)

\subsection{Comparison between [OIII] line width and $\sigma_{\star}$}
\begin{figure*}
  \includegraphics[scale=0.42]{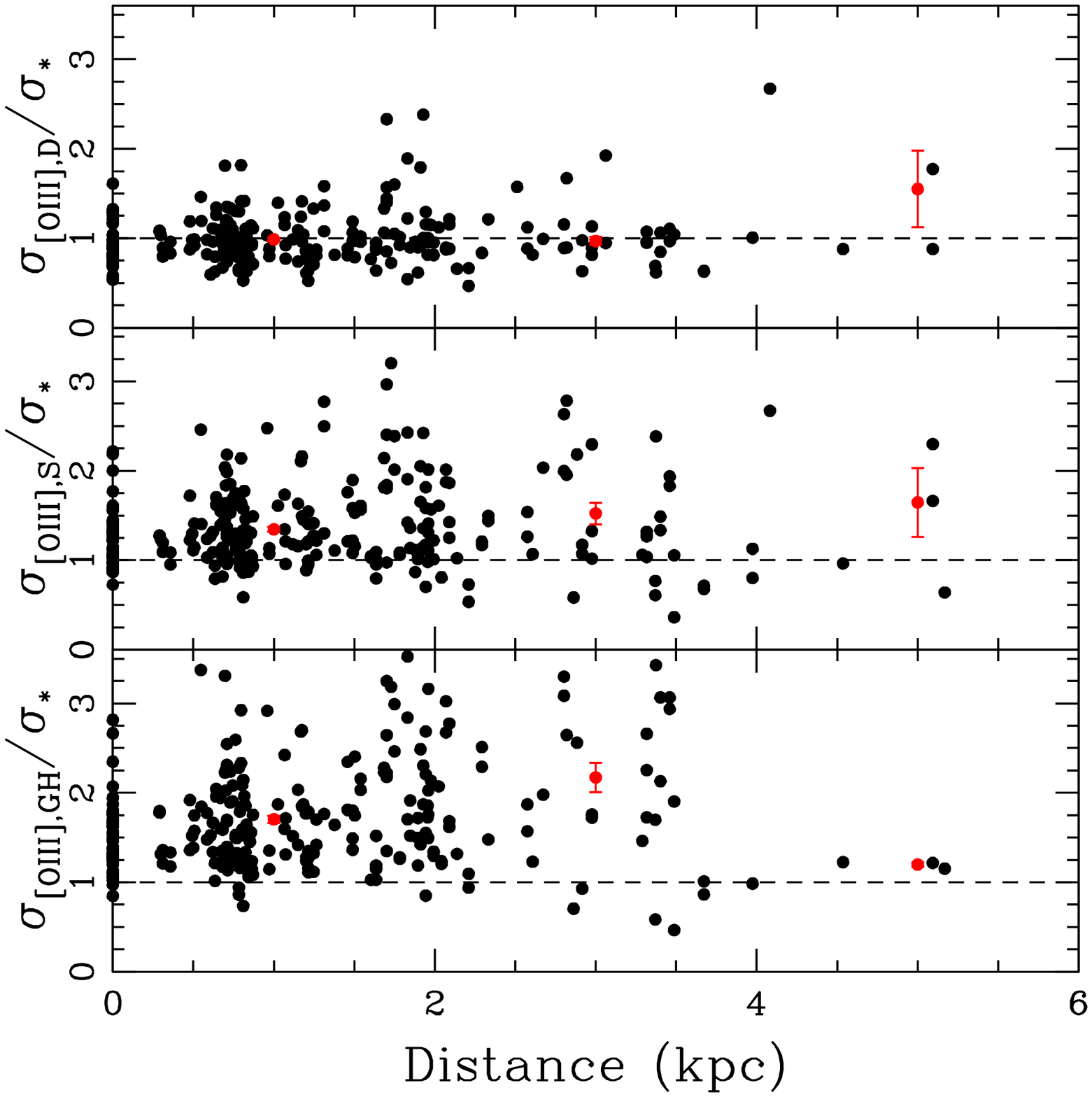}
  \includegraphics[scale=0.42]{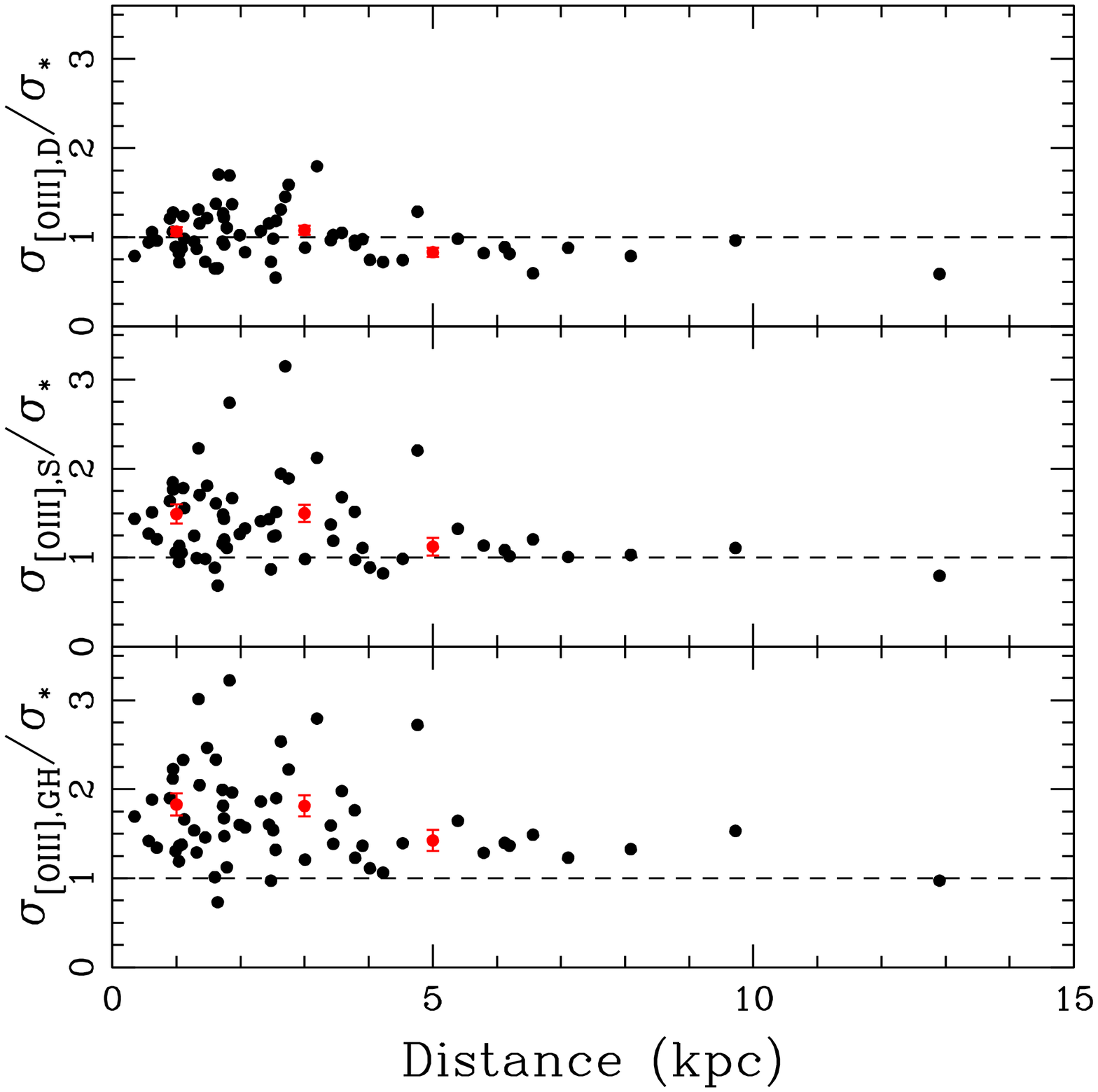}
  \caption{
    Ratio of [OIII] width ($\sigma_{\rm [OIII]}$)
    to stellar-velocity dispersion ($\sigma_{\star}$) as function of distance from galaxy center. Left:
    for spatially-resolved spectra.
    Right: for aperture spectra integrated
    over effective bulge radius.  Red data points show average ratio within
    distance bins 0-2 kpc, 2-4 kpc and 4-6 kpc (4-10 kpc for right panel), respectively.}
\label{figure:ratiodistance}
\end{figure*}

We compare the [OIII] line width ($\sigma_{\rm [OIII]}$) derived
from the three different fitting methods
(single Gaussian, double Gaussian using the central core component only, and Gauss-Hermite polynomials)
with the stellar-velocity dispersion ($\sigma_{\star}$).
In Figure~\ref{figure:ratiodistance},
the resulting $\sigma_{\rm [OIII]}$/$\sigma_{\star}$ ratio is shown as a function of
distance from the center for
both the spatially resolved spectra (left panels) as well as the aperture spectra within
the bulge effective radius (right panels).
In summary, the results show that
both the single Gaussian fit as well as the Gauss-Hermite polynomial fit
result in an overestimation of $\sigma_{\star}$ by on average 50-100\%
(see Table~\ref{table:sigmacompare}).
In other words, the entire [OIII] line is broader by $\sim$75\% compared
to $\sigma_{\star}$.
However, when line asymmetries are fitted by a second Gaussian
and excluded, then the central core [OIII] emission-line width is a good
tracer of $\sigma_{\star}$ (mean ratio 1.06$\pm$0.02 for spatially-resolved spectra;
mean ratio 1.02$\pm$0.04 for spectra within aperture of effective radius
\footnote{Note that we list the standard deviation of the mean.}),
but with individual data points off by up to a factor of two.

Another approach to exclude line asymmetries would be to consider only the
  width (i.e., $\sigma$) of the first pure Gaussian term in the Gauss-Hermite polynomial fit.
  Note that the first term (an original symmetric Gaussian)
  can represent most of the core of the line profile,
  while the rest of the series (Gaussian multiplied by Hermite polynomials)
  represents deviations to better describe the observed data profile.
  The resulting mean ratio with $\sigma_{\star}$ is then reduced to 1.25$\pm$0.04.
  While this is significantly lower
  than using the width (i.e., line dispersion)
  of the full profile of the fit, it still overestimates $\sigma_{\star}$
by $\sim$25\%. This is likely 
due to the fact that the higher order series terms
can have negative values which might then be compensated
for by the Gaussian, resulting in an overestimation of the width by the Gaussian component
\citep[see also,][]{woo18}.

Within the uncertainties, our data do not show a strong dependency of 
the $\sigma_{\rm [OIII]}$/$\sigma_{\star}$ 
on distance from the galactic center for any of the three fitting methods.
At first sight, this might indicate
that the influence of outflows is not necessarily more dominant
in the central regions.
However, given the S/N ratio, we do not probe regions
outside the central few kpc.
Moreover, given the ground-based seeing of $\sim$1-1.5$\arcsec$
of these Keck long-slit spectra and given the redshift range of our sample,
the central 1-2 kpc are essentially unresolved (as mentioned above).

We probe the dependency of the
$\sigma_{\rm [OIII]}$/$\sigma_{\star}$ ratio on
the velocity offset of the second Gaussian, the wing component,
with respect to the central core Gaussian component (using the spatially resolved data).
For the majority of the objects and rows,
the [OIII] profile has a blue wing
(see previous section).
Fitting this wing with a separate Gaussian
results in $\sigma_{\rm [OIII],D}$/$\sigma_{\star}$ = 1.06$\pm$0.02.
For objects/rows with a red wing,
the core component $\sigma$ ratio is 
$\sigma_{\rm [OIII],D}$/$\sigma_{\star}$ = 1.01$\pm$0.03.
For objects/rows for which the second Gaussian component
fitted a broader underlying central component,
$\sigma_{\rm [OIII],D}$/$\sigma_{\star}$ = 1.05$\pm$0.06.
However, in all three cases, if these  non-gravitational kinematic (blueshifted/redshifted/broad central)
components are not excluded from the fit by a second Gaussian,
they result in an overestimation of $\sigma_{\star}$.
For a single Gaussian fit, $\sigma_{\star}$ is overestimated
by 50$\pm$4\% for blueshifted wings,
by 45$\pm$7\% for redshifted wings,
and by 49$\pm$8\% for central broadening.
A Gauss-Hermite Polynomial leads to an overestimation of 92$\pm$5\%
for blue wings, 94$\pm$10\% for red wings and 82$\pm$12\% for broader central components.
This shows the necessity of fitting a double Gaussian for all types of [OIII] profiles
(blue wing, red wing or broader center)
and considering only the narrow core component as a surrogate for $\sigma_{\star}$.

We also checked for dependencies of the 
$\sigma_{\rm [OIII]}$/$\sigma_{\star}$ ratio on
the velocity shift of the entire [OIII] profile
compared to the H$\beta$ absorption line from stars.
The only noticeable trend is that a handful of objects/rows with 
large $\sigma_{\rm [OIII]}$/$\sigma_{\star}$ ratio in the core [OIII]
(as fitted by the double Gaussian)
are among those with large blueshifted [OIII] lines with an offset of at least
-150\,km\,s$^{-1}$.
However, while [OIII] can be offset by
-300\,km\,s$^{-1}$ to 200\,km\,s$^{-1}$,
there is no strong trend between the
velocity shift and the
$\sigma_{\rm [OIII]}$/$\sigma_{\star}$ ratio,
regardless of fitting method.

The width of the [OIII] wing
(when fitted by a double Gaussian)
 is larger by an average factor of 2.95$\pm$0.06 compared to the
[OIII] core, without showing a trend
with distance from the center or
overall velocity shift of the [OIII]
line with respect to the H$\beta$ absorption line.
This result is consistent with \citet{woo16} for a sample
of $\sim$39,000 type-2 AGNs from SDSS.

To look for a possible physical origin of the scatter,
we test dependencies of the $\sigma_{\rm [OIII]}$/$\sigma_{\star}$
ratio on other AGN and host-galaxy parameters,
taken from our previous publications \citep{ben15,run16}.
In particular, we probe the relationship between the
$\sigma_{\rm [OIII]}$/$\sigma_{\star}$ ratio and BH mass, as well as
$L_{5100}$~luminosity, but do not find a relationship.
Likewise, there is no correlation between the $\sigma_{\rm [OIII]}$/$\sigma_{\star}$ ratio
and the [OIII]/H$\beta_{\rm narrow}$ flux ratio,
host-galaxy morphology, or host-galaxy inclination.
This is in line with results by \citet{ric06} who also did not find
any trends in residuals when compared to host galaxy and nuclear properties.
While our sample consists of radio-quiet objects,
we discuss the effect of radio jets further below.

Note that while integral-field spectroscopic studies
have found increasing
evidence of galaxies
with kinematically de-coupled stellar and
gaseous components
with fractions as large as 
$\sim$30-40\% in
elliptical and lenticular galaxies
\citep[see e.g.,][and references therein]{sar06,dav11,bar15},
the larger survey of MaNGA
finds only 5\% of kinematically misaligned
galaxies \citep{jin16}.
Moreover, out of these, 90\% reside in
early-type galaxies.
Given our sample of pre-dominantly late-type
galaxies ($\sim$77\% with host galaxies classified
as Sa or later; \citealt{ben15}),
we expect a negligible fraction of kinematically de-coupled galaxies
in our sample.
Indeed, the overall gas rotation curve (as traced by [OIII]) matches
that of the stellar rotation curve \citep{har12}, with rotational velocities up
to $\sim$$\pm$250\,km\,s$^{-1}$.

\begin{table*}
  \caption{Ratios of [OIII] width to stellar-velocity dispersion depending on fitting method and distance from center.
Col. (1): Extraction of spectra.
Col. (2): Fitting method of [OIII] emission line.
Col. (3): Mean and uncertainty (of the mean) of the resulting ratio of [OIII] width ($\sigma_{\rm [OIII]}$) to stellar-velocity dispersion ($\sigma_{\star}$) for all measurements.
Col. (4): Same as Col. (3), but for distance from center of 0-2 kpc.
Col. (5): Same as Col. (3), but for distance from center of 2-4 kpc.
Col. (6): Same as Col. (3), but for distance from center of 4-6 kpc (4-10 kpc in case of reff).
  }
\label{table:sigmacompare}
\begin{tabular}{ccccccc}
\hline
  Spectrum & [OIII] Fit & Mean Ratio & Mean Ratio & Mean Ratio & Mean Ratio\\
  & & Total & Bin 1 & Bin 2 & Bin 3 \\
(1) & (2) & (3) & (4) & (5) & (6)\\
\hline
Spatially resolved      & Double Gaussian           & 1.06$\pm$0.02 & 1.06$\pm$0.02 & 1.03$\pm$0.04 & 1.4$\pm$0.3\\
                        & Single Gaussian           & 1.49$\pm$0.03 & 1.45$\pm$0.03 & 1.6$\pm$0.1 & 2.2$\pm$0.5\\
                        & Gauss-Hermite Polynomials & 1.95$\pm$0.05 & 1.85$\pm$0.04 & 2.2$\pm$0.1 & 2.8$\pm$1\\
Within effective radius & Double Gaussian           & 1.02$\pm$0.04 & 1.06$\pm$0.05 & 1.08$\pm$0.07 & 0.83$\pm$0.05\\
                        & Single Gaussian           & 1.42$\pm$0.07 & 1.5$\pm$0.1 & 1.5$\pm$0.1 & 1.1$\pm$0.1\\
                        & Gauss-Hermite Polynomials & 1.74$\pm$0.08 & 1.8$\pm$0.1 & 1.8$\pm$0.1 & 1.4$\pm$0.1\\
\hline
\end{tabular}
\end{table*}

\subsection{Including [OII] in the comparison}
\begin{figure}
\includegraphics[width=\columnwidth]{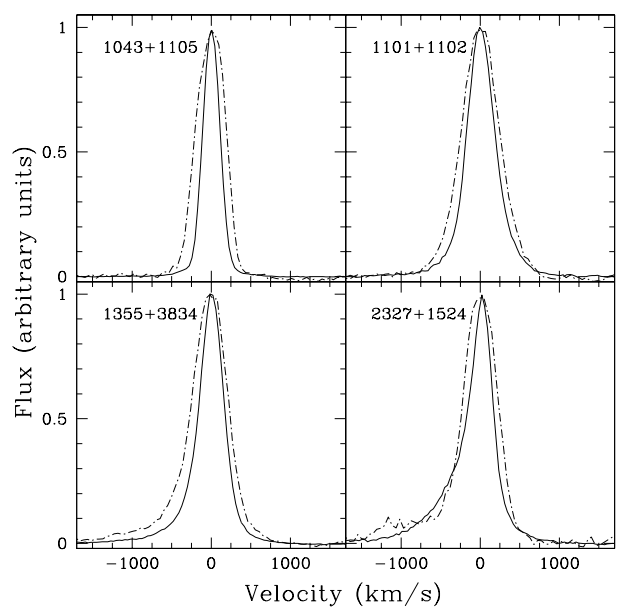}
  \caption{
      Examples of central [OIII] emission line (solid line) compared
      to [OII] (dash-dotted line).
      For comparison, the local continuum was
    subtracted and the peak flux scaled to 1.
     Since [OII] is a blended doublet line, it is broader than [OIII] in all cases.
      Blue wings seen in [OIII] are also present in [OII] (e.g., 1355+3834),
      but sometimes noisy, given the fainter [OII] line (e.g., 2327+1524).
       \label{figure:oiii_oii}}
\end{figure}

\begin{figure}
  \includegraphics[width=\columnwidth]{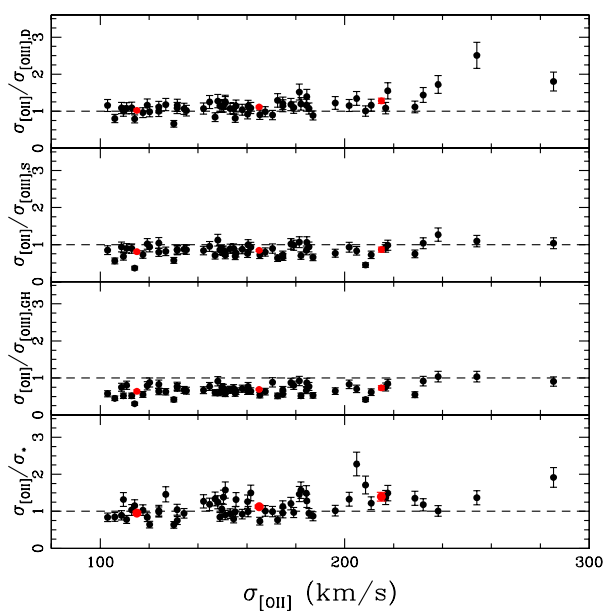}
  \caption{
    Ratio of [OII] width ($\sigma_{\rm [OII]}$)
    to [OIII] width as fitted with different methods ($\sigma_{\rm [OIII, D]}$, $\sigma_{\rm [OIII, S]}$,
    and $\sigma_{\rm [OIII, GH]}$) as well as to stellar-velocity dispersion ($\sigma_{\star}$; lower panel)
    as function of [OII] width ($\sigma_{\rm [OII]}$), for aperture spectra within the effective bulge radius.
Red data points show average ratio within
velocity bins
90-140 km\,s$^{-1}$,
140-190 km\,s$^{-1}$, and 240-290 km\,s$^{-1}$, respectively.
  }
\label{figure:oii_comparison}
\end{figure}

We compare the [OII] line width 
($\sigma_{\rm [OII]}$) with the 
[OIII] line width ($\sigma_{\rm [OIII]}$) derived
from the three different fitting methods
(single Gaussian, double Gaussian using the central component only, and Gauss-Hermite polynomials)
and with the stellar-velocity dispersion ($\sigma_{\star}$),
in all cases as derived from spectra of the central row
or within the bulge effective radius
(since these are the only spectra with [OII] width measurements,
given the lower S/N of [OII]).
Figure~\ref{figure:oiii_oii}
shows examples of a direct comparison the [OIII] and [OII] profiles.
In Figure~\ref{figure:oii_comparison},
the resulting ratios are shown as a function of [OII] width
($\sigma_{\rm [OII]}$) for the aperture spectra within the bulge effective radius.
Table~\ref{table:oiicompare} summarizes the average ratios, both overall
as well as a function of [OII] width.
To summarize, the [OII] width is smaller than the entire [OIII] line
(as represented by fits using a single Gaussian or Gauss-Hermite polynomials),
since the [OIII] line has prominent blue and red wings.
When these wings are excluded in a double Gaussian fit
and when comparing the narrow core component of [OIII] with [OII],
the widths are more comparable, but  the [OII] line
is  broader (on average by 17\%).
This can be attributed to wings that also appear in the [OII]
emission line, especially for larger widths:
while for 90\,km\,s$^{-1}$ $<$ $\sigma_{\rm [OII]}$ $<$ 140\,km\,s$^{-1}$,
the average ratio is 1.02$\pm$0.03,
the [OII] is wider by 12\% for velocities
140\,km\,s$^{-1}$ $<$ $\sigma_{\rm [OII]}$ $<$ 190\,km\,s$^{-1}$
and even up to 28\% wider for velocities
190\,km\,s$^{-1}$ $<$ $\sigma_{\rm [OII]}$ $<$ 240\,km\,s$^{-1}$.
This shows that while the lower ionization line has generally less prominent
wings from outflows (or infalls), they
are nevertheless present, especially for wider lines.
The same trend is observed when comparing 
$\sigma_{\rm [OII]}$ and $\sigma_{\star}$.
It is thus recommended to also fit the [OII] emission line
with a double Gaussian to exclude inflows and outflows as well,
i.e., using the same strategy as for the [OIII] fitting.
However, given that [OII] is already a blended doublet line,
the fitting of a double Gaussian to each individual line
is difficult, especially with low spectral resolution and S/N
which can often lead to the fitting of noise in the spectrum instead,
as our data showed.
Thus, using [OIII] is the better choice between both lines.
Our comparison cautions the use of low S/N emission lines (or spectra)
such as [OII] for which
the fitting of wings is more challenging.
Note that the results for [OII] determined from the central spectra are within the uncertainties
of those within the bulge effective radius and thus not further discussed here.

While the [SII] emission lines have also been found
to be a good substitute for $\sigma_{\star}$ \citep{gre05, kom07},
our spectral range does not cover these lines and we cannot make a direct comparison.
However, we suspect that [SII], also a line with
a lower ionization potential (23 eV), will behave similarly to [OII].

\begin{table*}
  \caption{Ratios of [OII] width to [OIII] width and stellar-velocity dispersion depending on fitting method and [OII] width. Col. (1): Extraction of spectra.
Col. (2): Fitting method of [OIII] emission line.
Col. (3): Mean and uncertainty (of the mean) of the resulting ratio of [OII] ($\sigma_{\rm [OII]}$) width to
[OIII] width ($\sigma_{\rm [OIII]}$) and stellar-velocity dispersion ($\sigma_{\star}$) for all measurements.
Col. (4): Same as Col. (3), but for [OII] width between 90-140\,km\,s$^{-1}$.
Col. (5): Same as Col. (3), but for [OII] width between 140-190\,km\,s$^{-1}$.
Col. (6): Same as Col. (3), but for [OII] width between 190-240\,km\,s$^{-1}$.
}
  \label{table:oiicompare}
  \begin{tabular}{ccccccc}
   \hline
Spectrum & [OIII] Fit & Mean Ratio & Mean Ratio & Mean Ratio & Mean Ratio\\
  &  & Total & Bin 1 & Bin 2 & Bin 3 \\
(1) & (2) & (3) & (4) & (5) & (6)\\
\hline
Effective radius (aperture)
& Double Gaussian & 1.17$\pm$0.04 & 1.02$\pm$0.03 & 1.12$\pm$0.03 & 1.28$\pm$0.07\\
& Single Gaussian & 0.86$\pm$0.02 & 0.81$\pm$0.04 & 0.84$\pm$0.02 & 0.87$\pm$0.07\\
& Gauss-Hermite Polynomials & 0.70$\pm$0.02 & 0.64$\pm$0.04 & 0.69$\pm$0.02 & 0.73$\pm$0.06\\
& Stellar-Velocity-Dispersion  & 1.15$\pm$0.04 & 0.95$\pm$0.05 & 1.13$\pm$0.04 & 1.4$\pm$0.1\\
\hline
  \end{tabular}
\end{table*}

\subsection{Black Hole Mass - $\sigma_{\rm [OIII]}$ relation}
\begin{figure*}
  \includegraphics[scale=0.42]{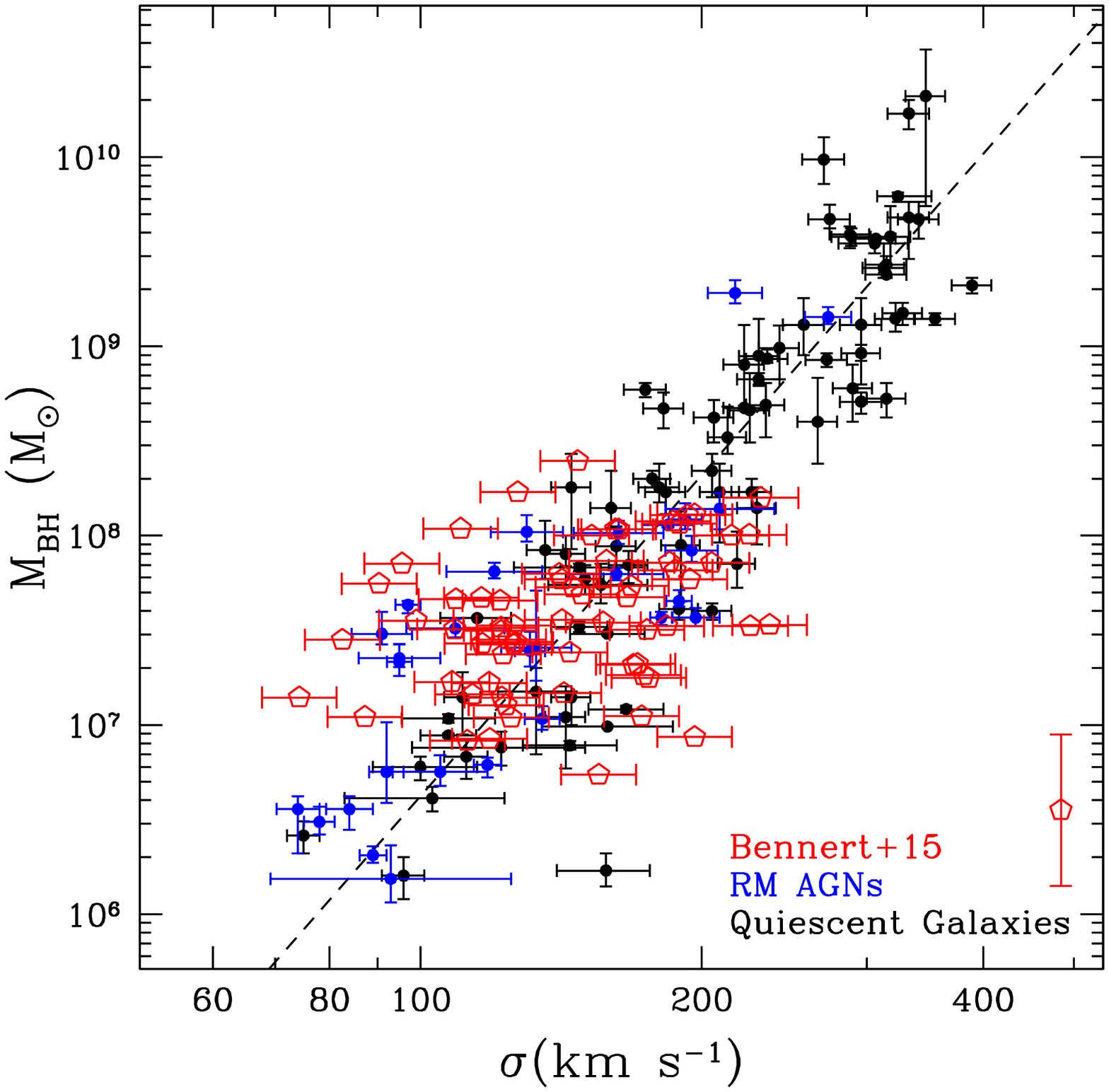}
\includegraphics[scale=0.42]{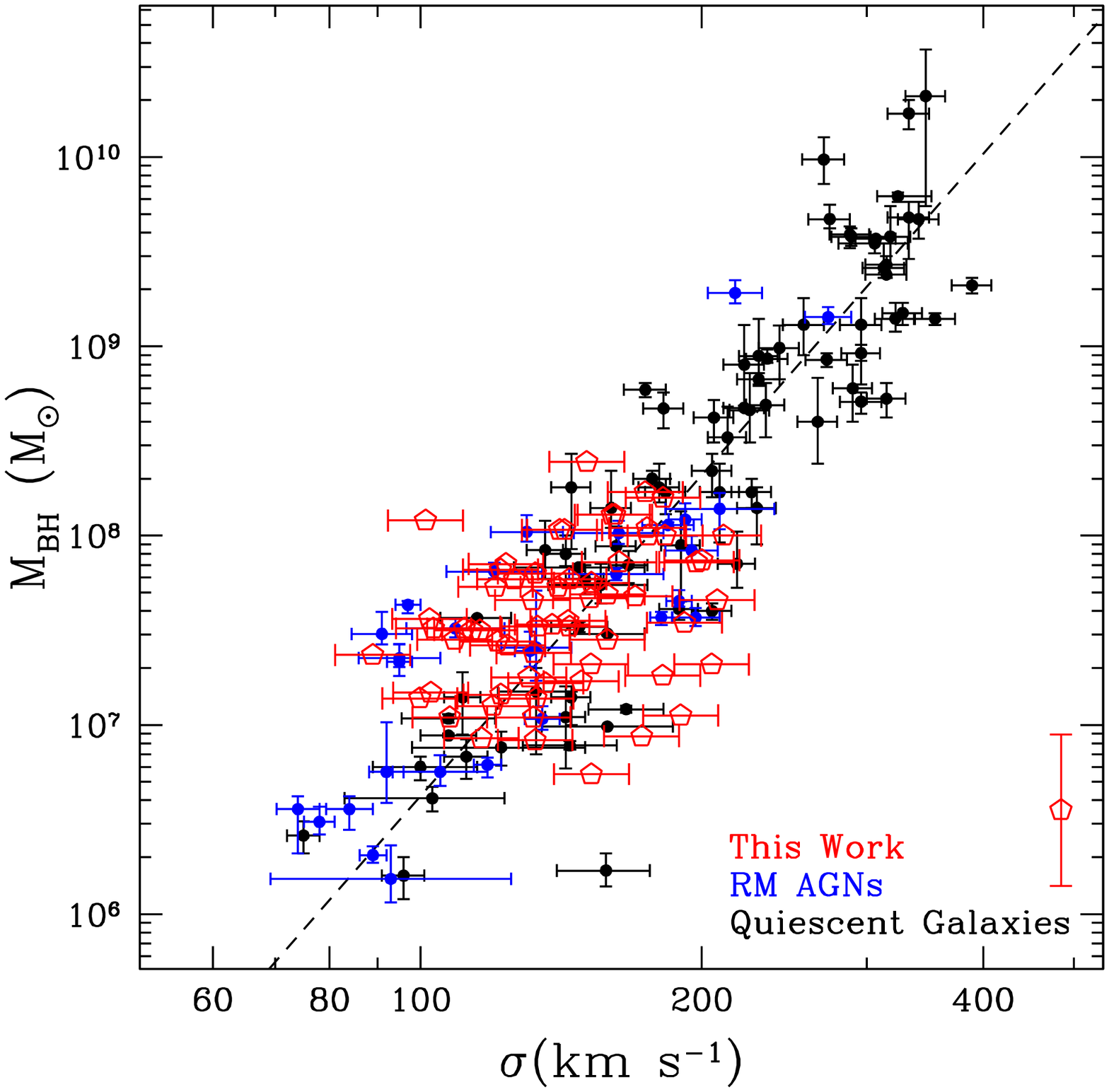}
  \includegraphics[scale=0.42]{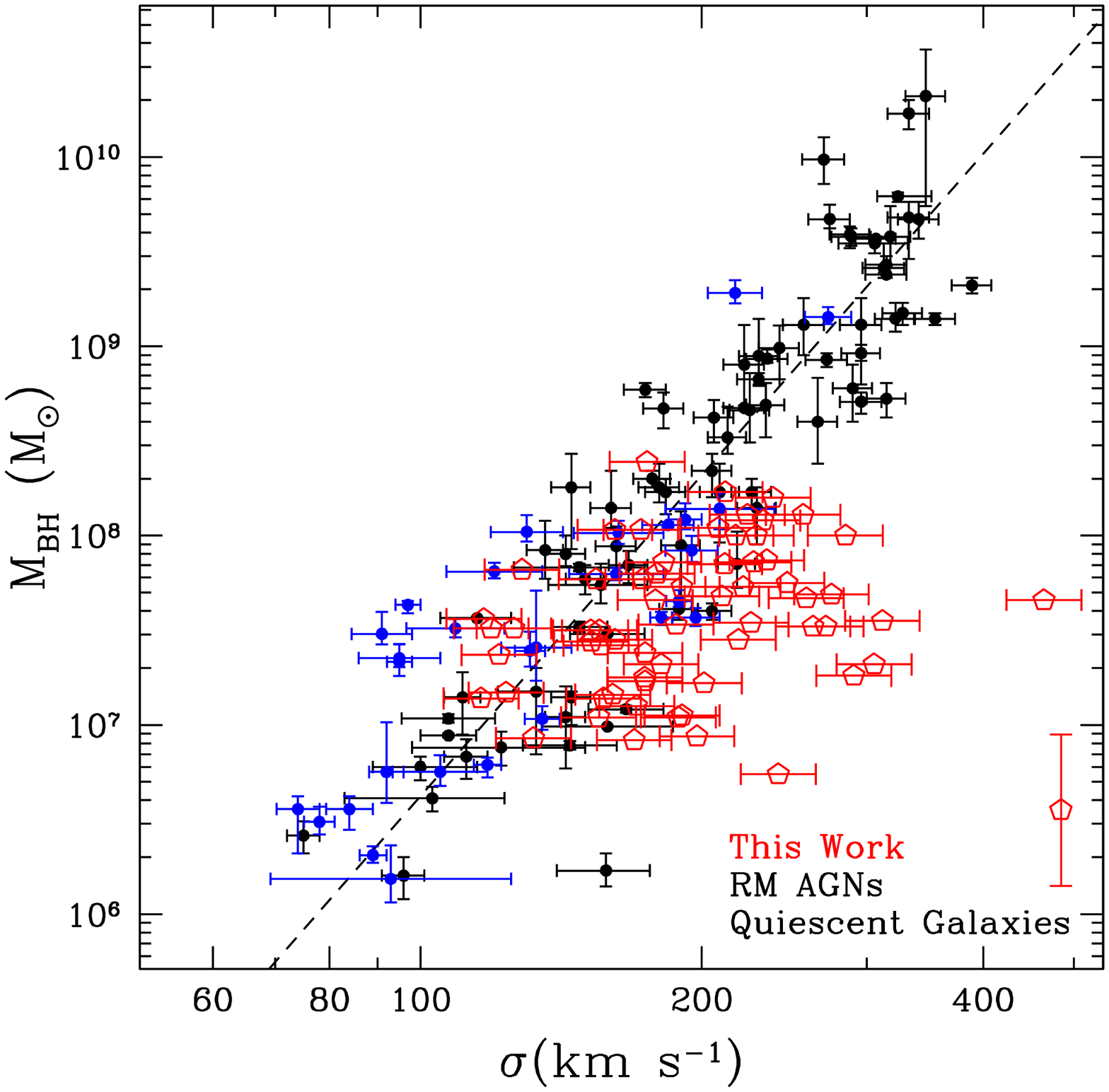}
  \includegraphics[scale=0.42]{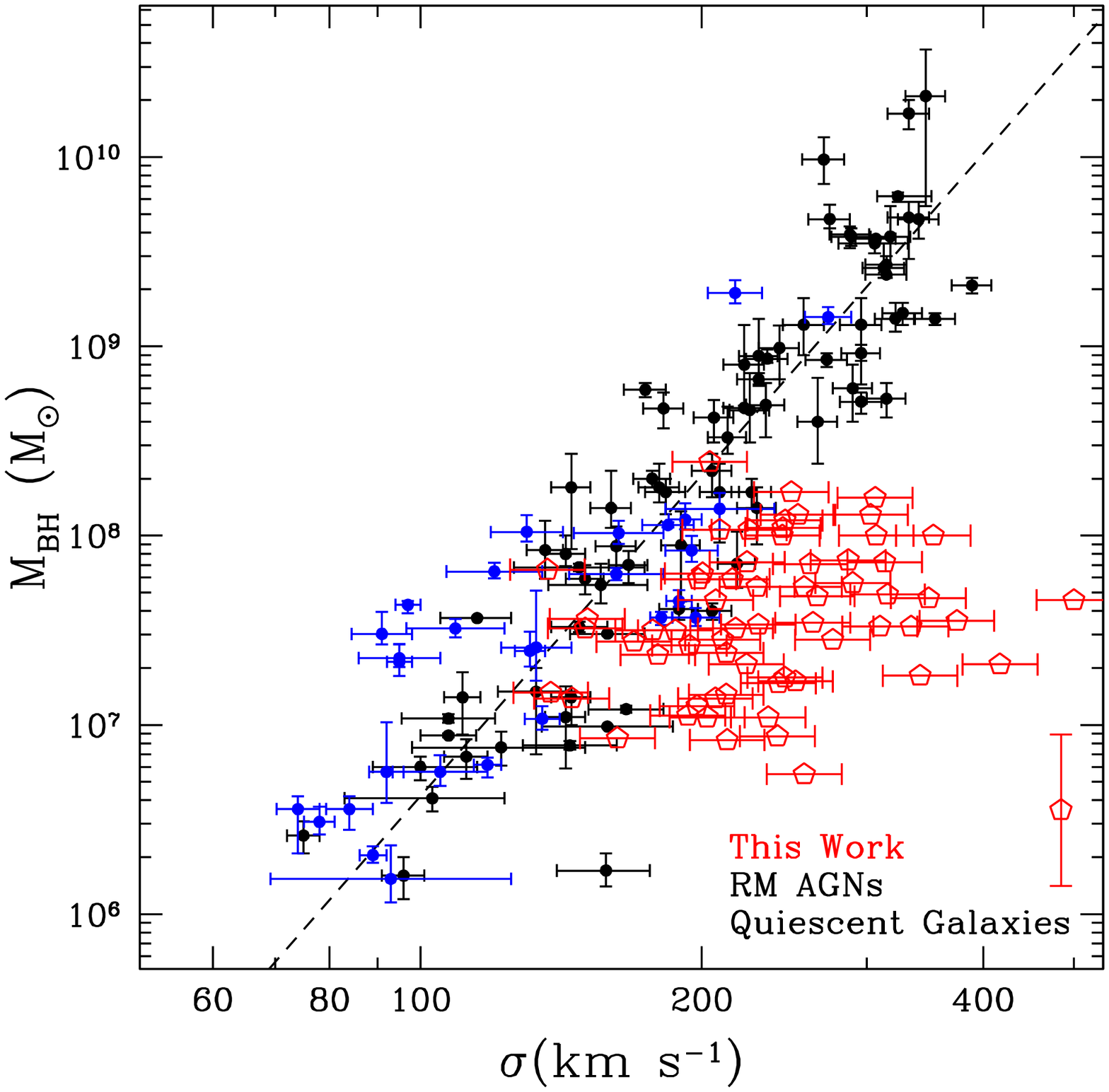}
\caption{
\mbh-\s~relation.
Upper left panel:
``True'' \mbh-\s~relation for 65 objects presented in \citet{ben15}
(red open pentagons),
reverberation-mapped AGNs \citep[blue;][]{woo15}, and a sample of quiescent local galaxies 
\citep[black;][with the black dashed line being their best fit]{mcc13}.
The error on the BH mass for our sample is 0.4 dex
and shown as a separate point with error bar in the legend, to reduce confusion of data points.
We assume a nominal uncertainty of the stellar-velocity dispersion of 0.04 dex.
Upper right panel:
The same as in the left panel, but for $\sigma_{\rm [OIII],D}$ (from aperture spectra
within effective bulge radius; our sample only) instead of $\sigma_{\star}$ (as shown in the left panel, also
derived within effective bulge radius).
Lower panels: The same as in the upper right panel, but using
  the [OIII] width as fitted by a single Gaussian (left panel) and
  Gauss-Hermite polynomials (right panel), in both cases clearly overestimating the ``true'' $\sigma_{\star}$.
}
\label{figure:mbhsigma}
\end{figure*}

We here compare the resulting \mbh-$\sigma_{\rm [OIII]}$ relations
 with the ``true''  \mbh-$\sigma_{\star}$ relation taken
from \citet{ben15}.
For comparison samples, we include quiescent galaxies \citep[][72 objects]{mcc13}
and reverberation-mapped AGNs \citep[][29 objects; adopting the same virial factor as
  for our sample; $\log f = 0.71$]{woo15}.
The results show that the total [OIII] emission line (as fitted by either
a single Gaussian and even more extreme for Gauss-Hermite Polynomials) overestimates $\sigma_{\star}$
and the points scatter to the right of the relation (Fig.~\ref{figure:mbhsigma}, bottom panels).
However, when based on $\sigma_{\rm [OIII],D}$, our sample follows the same \mbh-$\sigma_{\star}$ scaling relationship.

The systematic offset for the full [OIII] line width is significant,
especially since it is of the same order as the expected evolutionary
trend out to $z=1-2$ \citep[e.g.,][]{ben10,ben11b} and in the opposite direction.
In other words, using the width of the full [OIII] line as surrogate for $\sigma_{\star}$
(e.g., by simply fitting a single Gaussian)
in an attempt to study the evolution of the \mbh-\s~relation as done by
e.g., \citet[][]{sal13} will suggest a null result, even though there actually is significant
evolution.

Since our sample spans a small dynamical range in BH mass
(6.7$<$$\log$\mbh$<$8.2), 
and given the uncertainties of \mbh~of 0.4 dex,
we cannot determine the slope of the relationship independently.
Instead, we fit the data by the linear relation

\begin{eqnarray}
\log (M_{\rm BH}/M_{\odot}) = \alpha + \beta
\log(\sigma/200\,\rm km\,s^{-1})
\end{eqnarray}
taking into account uncertainties and
keeping the value of $\beta$ fixed to the corresponding relationships of
quiescent galaxies (5.64 for \cite{mcc13} and 4.38 for \citet{kor13})
or reverberation mapped AGNs \citep[][3.97]{woo15}.
The resulting zero point and scatter of the distribution
are comparable to that of the quiescent galaxies.
Table~\ref{fits_relations} summarizes the results for $\sigma_{\rm [OIII],D}$,
including a comparison to a quiescent galaxies sample taken from \citet[][51 objects; pseudo bulges and mergers excluded]{kor13}.

Note that the intrinsic scatter depends on the uncertainties of the measurements.
For the quiescent galaxy sample, \mbh~was derived from the kinematics of gas and/or stars
within the gravitational sphere of influence of the BH; for the comparison AGN sample,
\mbh~was derived more directly through reverberation mapping. Thus, for those samples, 
the uncertainty on \mbh~is significantly lower, on average 0.2dex and 0.15dex, respectively,
compared to 0.4dex for the single-epoch method used for our sample.
If, for example, for our \mbh-$\sigma_{\star}$,
we artificially assumed an uncertainty of \mbh~of 0.17dex, the scatter would increase
from  0.19 to 0.39
(for a fixed slope of 3.97), in other words, comparable to the 0.41 scatter of the reverberation-mapped AGN sample of
\citet{woo15}.
Thus, the most direct comparison of scatter is between the scatter
of the \mbh-\s~relation from \citet{ben15} and that of
the \mbh-$\sigma_{\rm [OIII],D}$ relation here,
since these are the identical samples with the same uncertainties in the \mbh ~measurements.
Independent of assumed fixed slope, we find a smaller scatter in
the \mbh-$\sigma_{\rm [OIII],D}$
This is likely due to the fact that
$\sigma_{\rm [OIII],D}$
covers a smaller dynamic range than $\sigma_{\star}$ (both within the effective bulge radius);
however, since the scatter is within the range of uncertainties, we do not discuss
this here further.

\begin{table*}
\caption{Fits to the local \mbh-$\sigma$ relation, $\log (M_{\rm BH}/M_{\odot}) = \alpha +
\beta \log (\sigma / 200 {\rm km\,s}^{-1})$.
Col. (1): Sample and sample size in parenthesis.
Col. (2): Mean and uncertainty on the best-fit intercept.
Col. (3): Mean and uncertainty on the best-fit slope.
Col. (4): Mean and uncertainty on the best-fit intrinsic scatter.
Col. (4): References for fit. 
Note that the quoted literature uses FITEXY with a uniform prior on the intrinsic scatter,
so our fits assume the same. The results from ``this paper'' are based on using $\sigma_{\rm [OIII],D}$ as surrogate
for $\sigma_{\star}$.
$^a$ Relation plotted as dashed lines in Fig.~\ref{figure:mbhsigma} and
used as fiducial relation when calculating residuals.}
\label{fits_relations}
\begin{tabular}{llcccc}
  \hline
Sample & $\alpha$ & $\beta$ & Scatter & Reference\\
(1) & (2) & (3)  & (4) & (5)\\
\hline
Quiescent Galaxies (72) & 8.32$\pm$0.05 & 5.64$\pm$0.32 & 0.38 & \citealt{mcc13}$^a$\\
Quiescent Galaxies (51)  & 8.49$\pm$0.05 & 4.38$\pm$0.29 & 0.29 & \citealt{kor13}\\
Reverberation-mapped AGNs (29)  & 8.16$\pm$0.18 & 3.97$\pm$0.56 & 0.41$\pm$0.05 & \citealt{woo15}\\
AGNs (66)              & 8.38$\pm$0.08 & 5.64 (fixed) & 0.43$\pm$0.09 &  \citealt{ben15}\\
AGNs (66)              & 8.20$\pm$0.06 & 4.38 (fixed) & 0.25$\pm$0.10 &  \citealt{ben15}\\
AGNs (66)              & 8.14$\pm$0.06 & 3.97 (fixed) & 0.19$\pm$0.10 &  \citealt{ben15}\\
\hline
AGNs (62)              & 8.41$\pm$0.07 & 5.64 (fixed) & 0.25$\pm$0.11 & this paper (based on $\sigma_{\rm [OIII],D}$)\\
AGNs (62)              & 8.23$\pm$0.06 & 4.38 (fixed) & 0.14$\pm$0.09 & this paper (based on $\sigma_{\rm [OIII],D}$)\\
AGNs (62)              & 8.16$\pm$0.06 & 3.97 (fixed) & 0.12$\pm$0.08 & this paper (based on $\sigma_{\rm [OIII],D}$)\\
\hline
\end{tabular}
\end{table*}

\subsection{Comparison with FIRST}
We searched the Very Large Array (VLA) Faint Images of the Radio Sky
at Twenty-Centimeters (FIRST) catalog for radio detection.
While our sample is radio quiet,
out of the 62 objects in the M$_{\rm{BH}}$-$\sigma_{\star}$ relation,
21 have been detected in FIRST, 37 objects have not been detected (FIRST detection limit $\sim$ 1mJy),
and 4 objects are outside of the survey area.
While for objects not detected in FIRST the ratio
$\sigma_{\rm [OIII],D}$/$\sigma_{\star}$ is comparable to the overall average of our sample,
i.e., close to 1,
(1.05$\pm$0.02 for spatially-resolved data and
0.99$\pm$0.04 for aperture spectra within the bulge-effective radius, respectively),
radio-detected objects have a larger width of [OIII],
overestimating $\sigma_{\star}$ by 13\% (the ratio is 1.13$\pm$0.03 for spatially-resolved
data and 1.13$\pm$0.06 for effective-radius integrated spectra).
When probing the broadening as a function of distance from the center, we see a trend
that it is more pronounced towards the nucleus.

\begin{figure*}
  \includegraphics[scale=0.42]{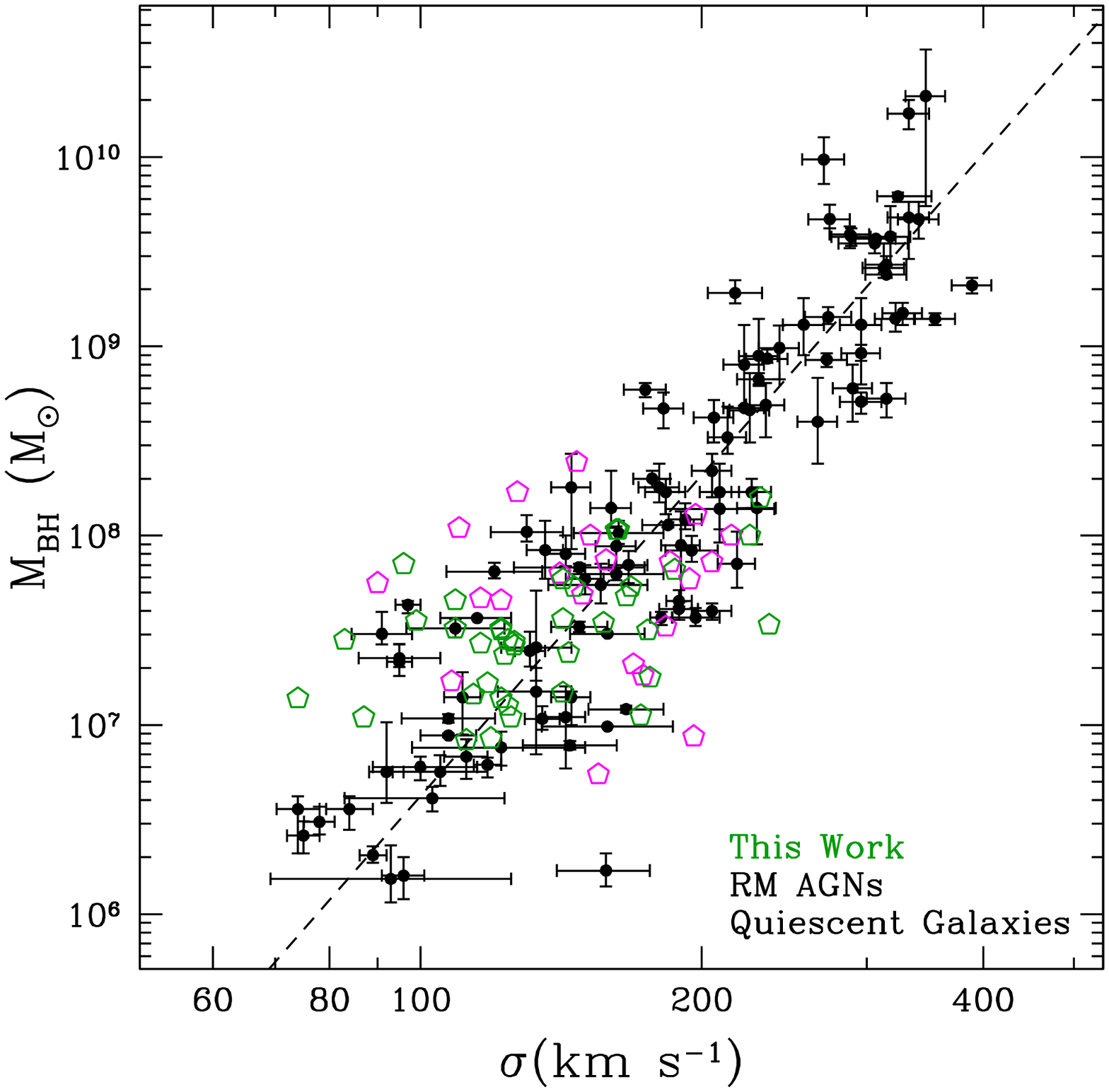}
\includegraphics[scale=0.42]{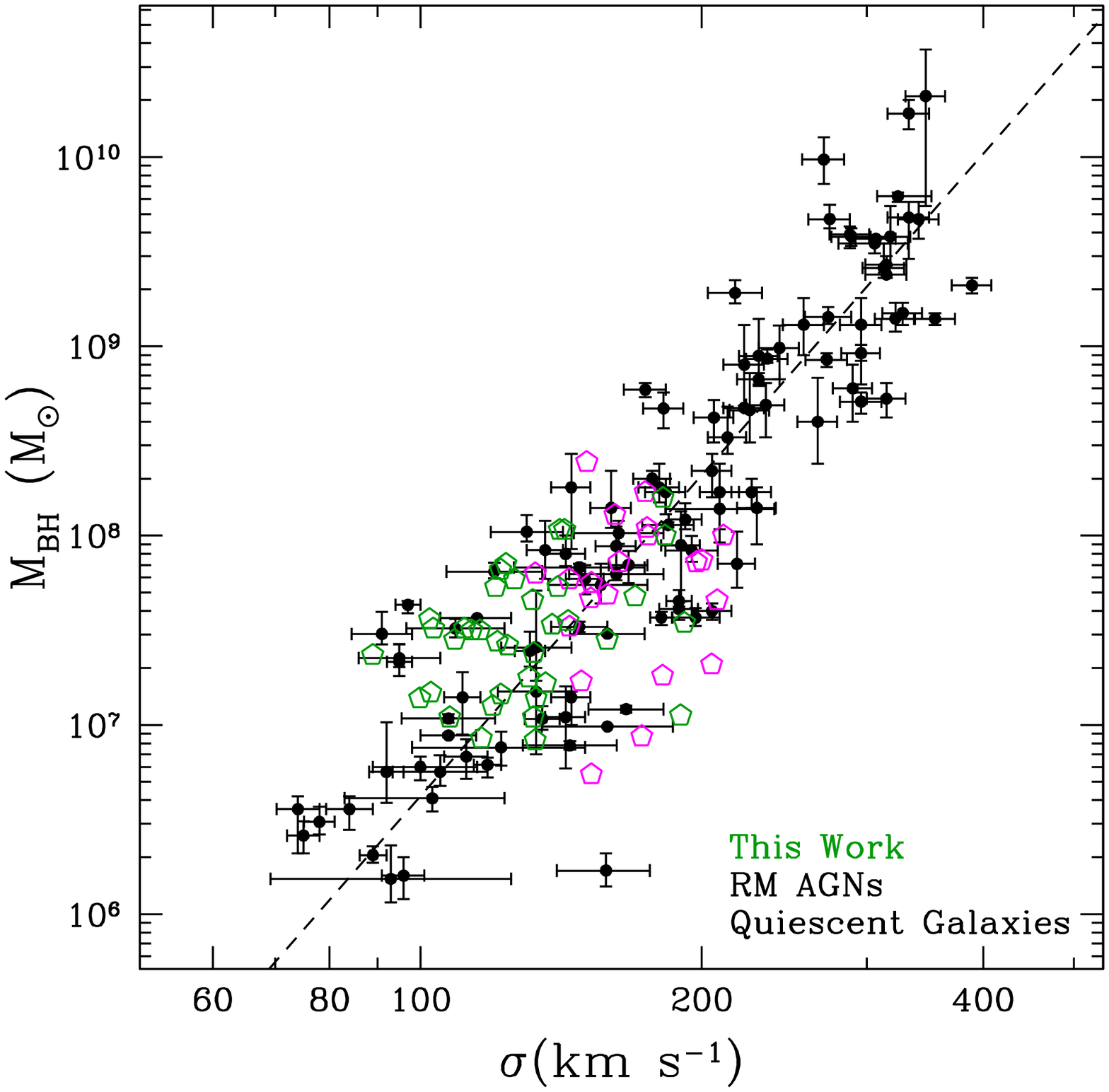}
\caption{
  Same as in Figure~\ref{figure:mbhsigma}, upper panels, but now distinguishing
  between objects detected in FIRST (magenta) and those with only upper limits (darkgreen);
  (literature samples shown in black). (No error bars shown to reduce confusion.)
  While there is no trend with $\sigma_{\star}$ (left panel), the radio does have a broadening effect
  on the [OIII] emission line (right panel), even when only considering the core of the line.
}
\label{figure:radiooiiiwidth}
\end{figure*}

We color-code objects accordingly in the
 \mbh-$\sigma$ relations (Figure~\ref{figure:radiooiiiwidth}).
In the M$_{\rm{BH}}$-$\sigma_{\star}$ relation, objects detected
in radio vs. those undetected by FIRST do not form distinct populations.
However, when using the width of the core [OIII] emission line (as traced by a double Gaussian, excluding
the wing component), there is a trend of objects detected in FIRST having larger
widths, especially those with lower \mbh.

Our results show that the radio emission, even in these radio-quiet objects,
has an effect on the [OIII] emission, broadening its dispersion, even for the core component.
This effect has also been observed in radio-loud emission-line galaxies, where the [OIII] central
component shows a strong trend of increasing line width with increasing central [OIII] peak shift
(i.e., outflow velocity), likely due to strong jet-cloud interactions across the NLR
\citep{kom18}.

\subsection{The potential and limitations of [OIII] width as a surrogate for $\sigma_{\star}$}
Overall, the results presented above are in agreement with those of previous studies,
concluding that the width of the narrow core
of the [OIII] emission line can be used as a replacement for $\sigma_{\star}$,
albeit with a large scatter \citep{nel00,gre05},
when considering only the central [OIII] component \citep{kom07,woo16},
when excluding sources with a blueshifted central [OIII] component
since these objects show strong additional line broadening \citep{kom08},
and when excluding objects with strong radio emission \citep{kom18}.
The resulting \mbh-$\sigma_{\rm [OIII],D}$ correlation scatters
around the known relation of quiescent galaxies.

\begin{figure*}
  \includegraphics[scale=0.42]{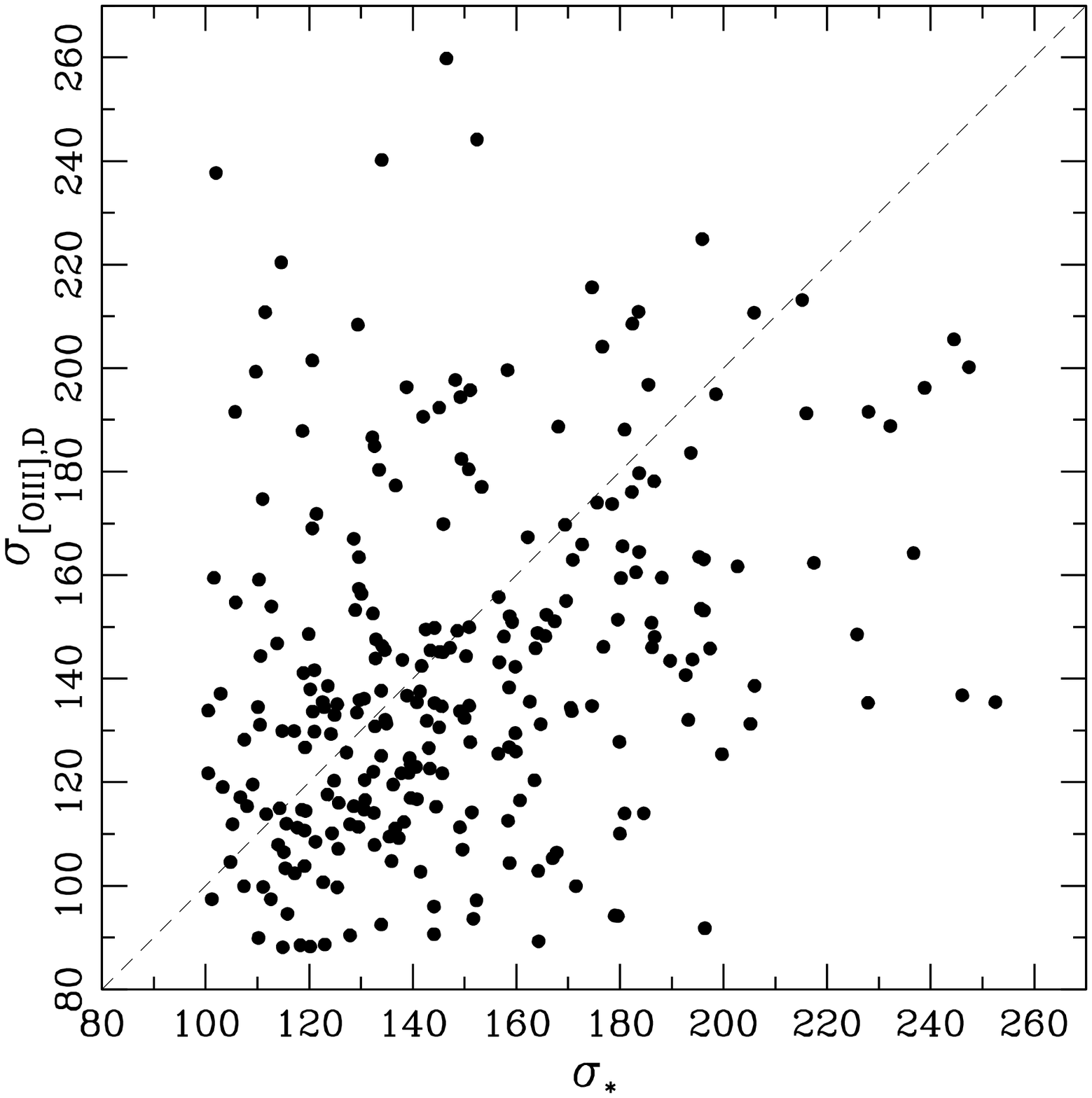}
  \includegraphics[scale=0.42]{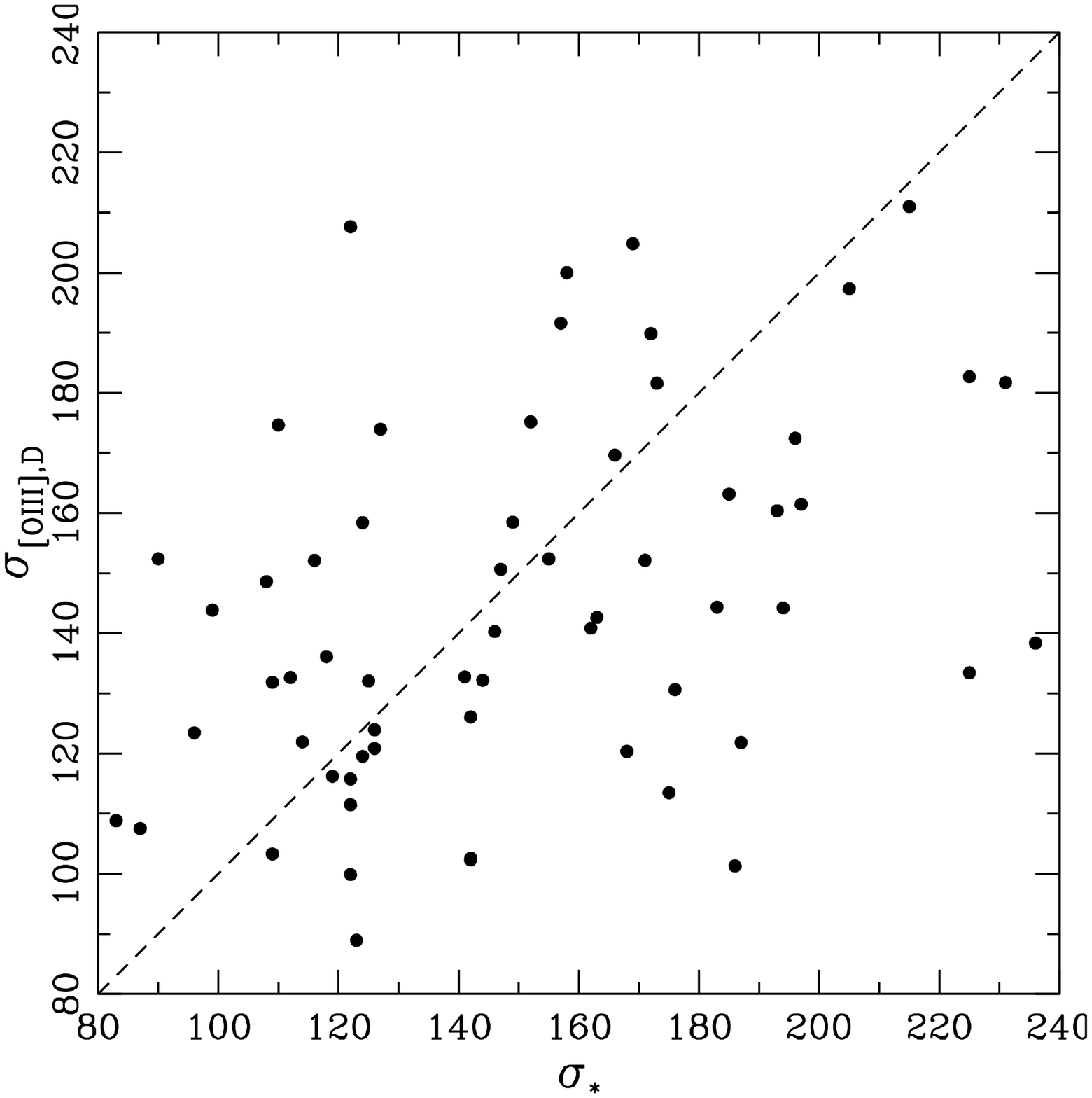}
\caption{
  Direct comparison between $\sigma_{\star}$ and $\sigma_{\rm [OIII],D}$. 
  The dashed line indicates the one-to-one relation.
  The left panel shows the result from spatially-resolved data, the right panel
  from aperture data, integrated within the effective bulge radius.
}
\label{figure:sigma_sigmaoiii}
\end{figure*}

However, when a direct comparison is made by plotting $\sigma_{\star}$
against $\sigma_{\rm [OIII],D}$,
either from spatially-resolved data or integrated within an aperture
of the effective bulge radius,
there is no strong correlation between the two (Figure~\ref{figure:sigma_sigmaoiii};
Pearson linear correlation coefficients of 0.25 for spatially-resolved data and 0.41 for aperture data; same results for Spearman rank correlation coefficient; see also \citet{liu09}).
This holds for both the radio-detected objects in the sample
as well as the ones not detected in FIRST.
Instead of a direct correlation between $\sigma_{\star}$
and $\sigma_{\rm [OIII],D}$, our data show that they cover the same range,
and that their average and standard deviation are similar.
Since we did not select on either quantity,
but purely on H$\beta$ width\footnote{Note that we
  are limited by our spectral resolution of 88\,km\,s$^{-1}$.},
this indicates a physical
connection and that they feel the same overall gravitational potential.
As a consequence, the ratio of $\sigma_{\rm [OIII],D}$ to $\sigma_{\star}$ is
close to one with a small deviation of the mean.
And since we start out with a \mbh-\s~relation that follows
that of quiescent galaxies and reverberation-mapped AGNs,
this naturally results in \mbh-$\sigma_{\rm [OIII],D}$ that scatter
around the same relation.
Given the large uncertainty in \mbh~based on single-epoch masses (0.4 dex),
a factor of  2 in $\sigma_{\rm [OIII],D}$/$\sigma_{\star}$ is still not that large.

At first sight, the absence of a strong correlation
could be due to the fact that we cover a relatively
small dynamic range in \mbh, especially given the large uncertainty in \mbh:
the range covered is roughly twice the uncertainty.
However, this is not true for measurements of $\sigma$:
For $\sigma_{\star}$, our sample has a factor of $\sim$3 in dynamic range
with a relatively small uncertainty (the range covered is roughly
seven times the uncertainty). Thus, the fact that we do not find a close correlation is significant.
While we cannot exclude that adding galaxies with larger $\sigma_{\star}$
would result in a trend, especially when considering mainly elliptical galaxies
for which the underlying kinematic field is simpler,
our sample consisting of AGNs hosted in mostly spiral galaxies
\citep[77\% classified as Sa or later;][]{ben15}
does not exhibit a significant correlation between
$\sigma_{\star}$ and $\sigma_{\rm [OIII],D}$.
8 objects have been conservatively classified as having a pseudo-bulge \citep{ben15}.
  These objects are not amongst any particular outliers in the $\sigma_{\star}$ and $\sigma_{\rm [OIII],D}$
  plots. However, the sample size is small and the classification based on SDSS images for which
  a morphological classification is difficult, given the presence of the bright AGN point source.
  We will re-visit the question of pseudo-bulges with higher-resolution images 
  (HST-GO-15215; PI: Bennert).

We consider the careful fitting of a double Gaussian, excluding wings and the use of the narrow
core component for estimation of the [OIII] width, a robust approach;
the $\sigma_{\star}$ measurements were taken with an equally great care \citep{har12}.
Our sample further has the advantage of high S/N spatially-resolved spectra,
allowing a direct comparison of  $\sigma_{\rm [OIII],D}$ and $\sigma_{\star}$  
for the same object, using the same spectra and the same aperture.
Thus, the reason for the scatter is likely a physical one.
Generally speaking, both absorption and emission line profiles are a luminosity-weighted line-of-sight average,
depending on the light distribution and the underlying kinematic field
which can be different between gas and stars.
Also, there may still be effects of outflows, inflows, and anisotropies 
not accounted for in the double Gaussian fitting of [OIII].
Finally, radiation pressure would only act on gas, not on stars.
Our data do not allow to single out any of these as the main cause of the scatter.

Given the high quality of our kinematic data, both in terms of S/N of the spectra
as well as the detailed fitting, the fact that we do not find a close correlation
between $\sigma_{\rm [OIII],D}$ and $\sigma_{\star}$
strongly cautions against the use of  $\sigma_{\rm [OIII],D}$ as a surrogate for $\sigma_{\star}$
on an individual basis, even though as an ensemble they trace the same gravitational potential.

\section{Summary}
\label{summary}
We study the spatially-resolved [OIII]$\lambda$5007\AA~emission line profile
of a sample of $\sim$80 local ($0.02 \leq z \leq 0.1$) type-1 Seyfert galaxies.
Stellar-velocity dispersion (\s) derived from high-quality long-slit Keck spectra
  is used to probe whether the width of the [OIII] emission line,
obtained by three different methods, is a valid substitute for \s.
Since the [OIII] emission line is known to often have broad wings
from non-gravitational motion, such as outflow, infall or turbulence,
we fit the line with a double Gaussian.
For comparison, we include a single Gaussian fit, the simplest fit,
and Gauss-Hermite polynomials 
which yield the best overall fit to the line.
Our results can be summarized as follows.

\begin{enumerate}

\item
In 66\% of the spectra we find the presence of a blue wing,
22\% of the spectra show a red wing,
and in 12\% of the cases, a broader central component is seen.

\item
The width of the narrow core component of [OIII]
from a double Gaussian fit,
is, on average, the closest tracer of \s~(mean
ratio of 1.06$\pm$0.02 for spatially-resolved
spectra and 1.02$\pm$0.04 for spectra within aperture of
effective radius, respectively).
However, the scatter is large, with individual objects
off by up to a factor of 2.

\item
Fitting [OIII] with a single Gaussian or 
Gauss-Hermite polynomials results in a width that is, on average, 50-100\%
larger than the stellar-velocity dispersion.
This strongly cautions against the use of the full [OIII] width
as a surrogate for $\sigma_{\star}$ in evolutionary studies,
since the systematic offset will mimic a null-result.

\item
We do not find trends of the $\sigma_{\rm [OIII]}/\sigma_{\star}$ ratio
with distance from the center nor dependencies on other properties
of the AGN (such as BH mass and $L_{5100}$) or the host galaxy (such as morphology and inclination).

\item
Even though our sample consists of radio-quiet Seyfert galaxies,
$\sim$30\% have FIRST detections.
The radio emission even effects the [OIII] core width,
leading to $\sim$10\% broader lines.

\item
When considering the width of the narrow core component of [OIII],
the resulting \mbh-$\sigma_{\rm [OIII],D}$
relation scatters around the \mbh-\s~relations
of quiescent galaxies and reverberation-mapped AGNs.

\item
We compare the width of the doublet [OII]$\lambda\lambda$3726,3729\AA\AA,
fitted by a double Gaussian, with those of [OIII] and \s.
While wings are less prominent in the low-ionization [OII] line, they are nevertheless
present, especially for wider lines, but harder to fit given the blended nature of the
line and the lower S/N. Thus, [OIII] is preferable over [OII].

\item
A direct comparison between $\sigma_{\star}$ and
$\sigma_{\rm {[OIII]}}$ shows that there is no correlation on an individual basis.
Overall, gas and stars follow the same gravitational potential
and thus have similar distributions 
in terms of range, average, and standard deviation.
This results in an average ratio of $\sigma_{\rm [OIII],D}$ to $\sigma_{\star}$
close to one, with a small deviation of the mean,
and an \mbh-$\sigma_{\rm [OIII],D}$ relation that scatters
around those of quiescent galaxies and reverberation-mapped AGNs.

\item
  The reason for the large scatter is likely a physical one.
  Line profiles are luminosity-weighted line-of-sight averages
  that depend  on the light distribution and the underlying kinematic field
  which can be different between gas and stars.
  Moreover, effects of outflows, inflows, anisotropies and radiation pressure
  on the [OIII] profile not accounted for in the double Gaussian fitting
   can increase the scatter.
 
\item
Given the large dynamic range covered in $\sigma_{\star}$
and the high quality of our kinematic data,
our results are significant and caution against the use
of [OIII] as a surrogate for $\sigma_{\star}$ on a case-by-case basis,
even though as an ensemble they trace the same gravitational potential.

\end{enumerate}

\section*{Acknowledgements}
VNB thanks Dr. Bernd Husemann and Dr. Knud Jahnke for discussions.
VNB is grateful to Dr. Hans-Walter Rix and the Max-Planck Institute
for Astronomy, Heidelberg, for the hospitality and financial support during her sabbatical stay.
VNB, DL, ED, MC, SL, and NM gratefully acknowledge
assistance from a National Science Foundation (NSF) Research at Undergraduate Institutions (RUI) grant AST-1312296.
Note that findings and conclusions do not necessarily represent views of the NSF.
This research has made use of the Dirac computer cluster at Cal Poly, maintained by Dr. Brian Granger and Dr. Ashley Ringer McDonald.  Spectra were obtained at the W. M. Keck Observatory, which is operated as a scientific partnership among Caltech, the University of California, and NASA.  The Observatory was made possible by the generous financial support of the W. M. Keck Foundation.  The authors recognize and acknowledge the very significant cultural role and reverence that the summit of Mauna Kea has always had within the indigenous Hawaiian community.  We are most fortunate to have the opportunity to conduct observations from this mountain.  This research has made use of the public archive of the Sloan Digital Sky Survey (SDSS) and the NASA/IPAC Extragalactic Database (NED), which is operated by the Jet Propulsion Laboratory, California Institute of Technology, under contract with the National Aeronautics and Space Administration.

\bsp
\label{lastpage}
\end{document}